\documentclass[fleqn,usenatbib]{mnras}

\usepackage{newtxtext,newtxmath}

\usepackage[T1]{fontenc}

\DeclareRobustCommand{\VAN}[3]{#2}
\let\VANthebibliography\thebibliography
\def\thebibliography{\DeclareRobustCommand{\VAN}[3]{##3}\VANthebibliography}

\usepackage{graphicx}	
\usepackage{amsmath}	
\usepackage{subcaption}

\newcommand{\hess}{H.\,E.\,S.\,S.\,}

\title[Non-thermal Emission from CXOU J1714-3810]{Non-thermal emission from the vicinity of the magnetar CXOU J171405.7-381031}

\author[M. F. Sousa and R. C. Anjos]{
Manoel F. Sousa,$^{1}$\thanks{E-mail: mfelipe.sousa@ifsc.usp.br}
Rita. C. Anjos,$^{2,3,4,5,6,7}$
\\
$^{1}$Instituto de Física de São Carlos, Universidade de São Paulo, Av. Trabalhador São-carlense 400, São Carlos, Brasil\\
$^{2}$Centro de Artes, Humanidades e Tecnologia, Universidade Federal de São Carlos, Eduardo Nielsen, 420, 15030-070 São José do Rio Preto, SP, Brazil\\
$^{3}$Departamento de Engenharias e Exatas, Universidade Federal do Paraná, Pioneiro, 2153, 85950-000 Palotina, PR, Brazil\\
$^{4}$Programa de Pós-Graduação em Física e Astronomia, Universidade Tecnológica Federal do Paraná, 80230-901 Curitiba, PR , Brazil\\
$^{5}$N\'ucleo de Astrof\'{\i}sica e Cosmologia, Universidade Federal do Esp\'irito Santo, 29075-910 Vit\'oria, ES, Brazil\\
$^{6}$Programa de pós-graduação em Física, Universidade Estadual de Londrina, Rodovia Celso Garcia Cid Km 380, 86057-970 Londrina, PR, Brazil\\
$^{7}$Programa de Pós-Graduação em Física Aplicada, Universidade Federal da Integração Latino-Americana, 85867-670 Foz do Igua\c{c}u, PR, Brazil
}

\date{Accepted XXX. Received YYY; in original form ZZZ}

\pubyear{\the\year{}}

\begin{document}
\label{firstpage}
\pagerange{\pageref{firstpage}--\pageref{lastpage}}
\maketitle

\begin{abstract}
Magnetars are neutron stars with ultra-strong magnetic fields ($B \sim 10^{14}$--$10^{15}$~G) and are promising candidates for high-energy particle acceleration. We present a multiwavelength analysis of the region surrounding CXOU~J171405.7--381031, a magnetar associated with the supernova remnant (SNR) CTB~37B. The broadband spectral energy distribution spanning radio to TeV energies is modeled using leptonic and lepto-hadronic scenarios, with particle populations constrained using Markov Chain Monte Carlo techniques. Within an SNR framework, both scenarios provide acceptable descriptions of the gamma-ray data. The purely leptonic model reproduces the overall spectral shape but slightly underestimates the highest-energy flux measured by H.E.S.S., whereas a lepto-hadronic interpretation offers an improved description above $\sim 10$~TeV, with inverse-Compton scattering dominating the GeV emission and neutral-pion decay contributing at the highest energies. The required proton energy ($W_{\rm p} \gtrsim 10^{51}$~erg) can be substantially reduced if the remnant interacts with a dense ambient medium. A magnetar wind nebula scenario can reproduce the broadband spectrum but is strongly disfavored by the observed source morphology. Simulated Cherenkov Telescope Array Observatory (CTAO) observations indicate that exposures of $\sim 50$~h will constrain the proton cut-off energy, enabling a decisive test of hadronic emission in this region.

\end{abstract}




\section{Introduction} \label{sec:intro}

High-energy astrophysical studies aim to identify the most efficient cosmic accelerators within and beyond our Galaxy \citep[see e.g.][]{2021JCAP...10..023D, Coelho_2022, 2023MNRAS.519..136V, 2023ARNPS..73..341C, 2024A&A...689A...9D, 2024PhRvL.132i1401P, 2025MNRAS.luana, 2025A&A...693A.255A}. Among the various candidates, magnetars, ultra-magnetized neutron stars with surface fields up to $10^{15}$~G, stand out as compelling sources of non-thermal radiation and energetic outflows that can power particle acceleration in their surrounding environments \citep{2010MNRAS.406L..25H, PhysRevD.84.023002, 2015RPPh...78k6901T, 2017ARA&A..55..261K, 2018KPCB...34..167G}. Recent multiwavelength observations, particularly in the gamma-ray domain, have highlighted the role of magnetars in contributing to the Galactic cosmic-ray population, bridging observational and theoretical gaps in our understanding of particle acceleration mechanisms \citep{2005Natur.434.1107P, 2009MNRAS.397.1420B, 2021ASSL..461...97E, 2025ApJ...979...23S, 2024MNRAS.531.3297K}.

A systematic investigation of gamma-ray emission from magnetar regions was conducted in the context of the upcoming Cherenkov Telescope Array Observatory (CTAO) by \citet{2025ApJ...979...23S}, following the methodology proposed by \citet{costa2024gamma}. Their study focused on three magnetar-associated regions -- CXOU~J171405.7--381031, Swift~J1834.9--0846, and SGR~1806--20 -- to assess their detectability with CTAO. The results demonstrated that these regions are promising gamma-ray sources, with expected CTAO detections reaching high significance levels, particularly for the CXOU~J171405.7--381031 (henceforth CXOU~J1714--3810) and Swift~J1834.9--0846 (henceforth Swift~J1834--0846) magnetar environments. However, while \citet{2025ApJ...979...23S} focused on observational detectability and spectral extrapolations, the physical mechanisms governing particle acceleration and emission in these environments remain uncertain.

Expanding on these preliminary findings, this paper presents a comprehensive theoretical investigation of particle acceleration processes in the region around a magnetar. We focus on one particularly promising region, CXOU~J1714--3810 \citep{2008A&A...486..829A, 2010PASJ...62L..33S, 1975A&A....45..239C}, employing detailed multiwavelength spectral modeling. In addition to the magnetar itself, this region hosts other potential particle accelerators and interaction sites capable of shaping the observed emission. Accordingly, our modeling accounts not only for the direct magnetar contribution but also for the surrounding environment: the supernova remnant (SNR) shell, a possible magnetar wind nebula (MWN), and potential SNR--molecular cloud (MC) interactions. This approach enables a self-consistent treatment of both leptonic and hadronic emission channels across the radio-to-TeV energy range \citep[e.g.][]{2009ASSL..357..421K, 2009ApJ...706L...1A, 2010ApJ...712..459A, 2010ApJ...718..348A}.

We model both purely leptonic and lepto-hadronic scenarios. Using Markov Chain Monte Carlo (MCMC) sampling within the \texttt{Naima} framework \citep{naima}, we fit the broadband spectral energy distribution (SED) and obtain posteriors for the particle populations and their radiative components (synchrotron, inverse-Compton (IC), and neutral-pion decay) following standard formalisms \citep[e.g.][]{1979rpa..book.....R, 1970PhRvD...1.1596B, 2006PhRvD..74c4018K}. This enables us to quantify each channel's contribution and assess whether the $\gamma$-ray emission is predominantly leptonic or hadronic \citep{Landshoff:1971xb, 2019scta.book.....C}. We also examine the dependence on environmental inputs (ambient gas density, magnetic field strength, and target photon fields), which govern acceleration efficiency and energy losses.

The results of this study have broad implications for understanding the role of magnetars and their wind nebulae as potential cosmic-ray accelerators. By characterizing the dominant energy-loss mechanisms in these regions and comparing them with known SNR--magnetar associations, we provide a refined picture of high-energy particle acceleration in compact object environments. Furthermore, our findings contribute to the growing anticipation for CTAO observations, offering theoretical predictions that can be tested against future high-sensitivity gamma-ray data.

Throughout this work, the terms ``magnetar region'' and ``region around the magnetar'' denote not only the immediate surroundings of the magnetar but also the broader environment that includes associated GeV--TeV sources, such as SNR shells, MWN, and nearby MC interactions. In these environments, particle acceleration is assumed to occur at the SNR forward shock or at the wind termination shock. Acceleration processes occurring directly within the magnetar magnetosphere are beyond the scope of the present modeling. This paper is structured as follows: Section~\ref{sec:magnetars} introduces the key properties and previous observational constraints of CXOU~J1714--3810. Section~\ref{sec:results_CXOU} examines the CXOU~J1714--3810 field (CTB~37B) within SNR and PWN scenarios, presenting multiwavelength spectral modeling under both leptonic and lepto-hadronic frameworks. In Section~\ref{sec:discussions_CXOU}, we discuss the broader implications of our results, while in Section~\ref{sec:CTAO} we present the prospects for our analysis, testing them with upcoming CTAO observations. Finally, Section~\ref{sec:conc} summarizes our main conclusions and outlines directions for future research.

\section{CXOU~J1714-3810 region}  \label{sec:magnetars}

\begin{figure}
    \includegraphics[width=\hsize,clip]{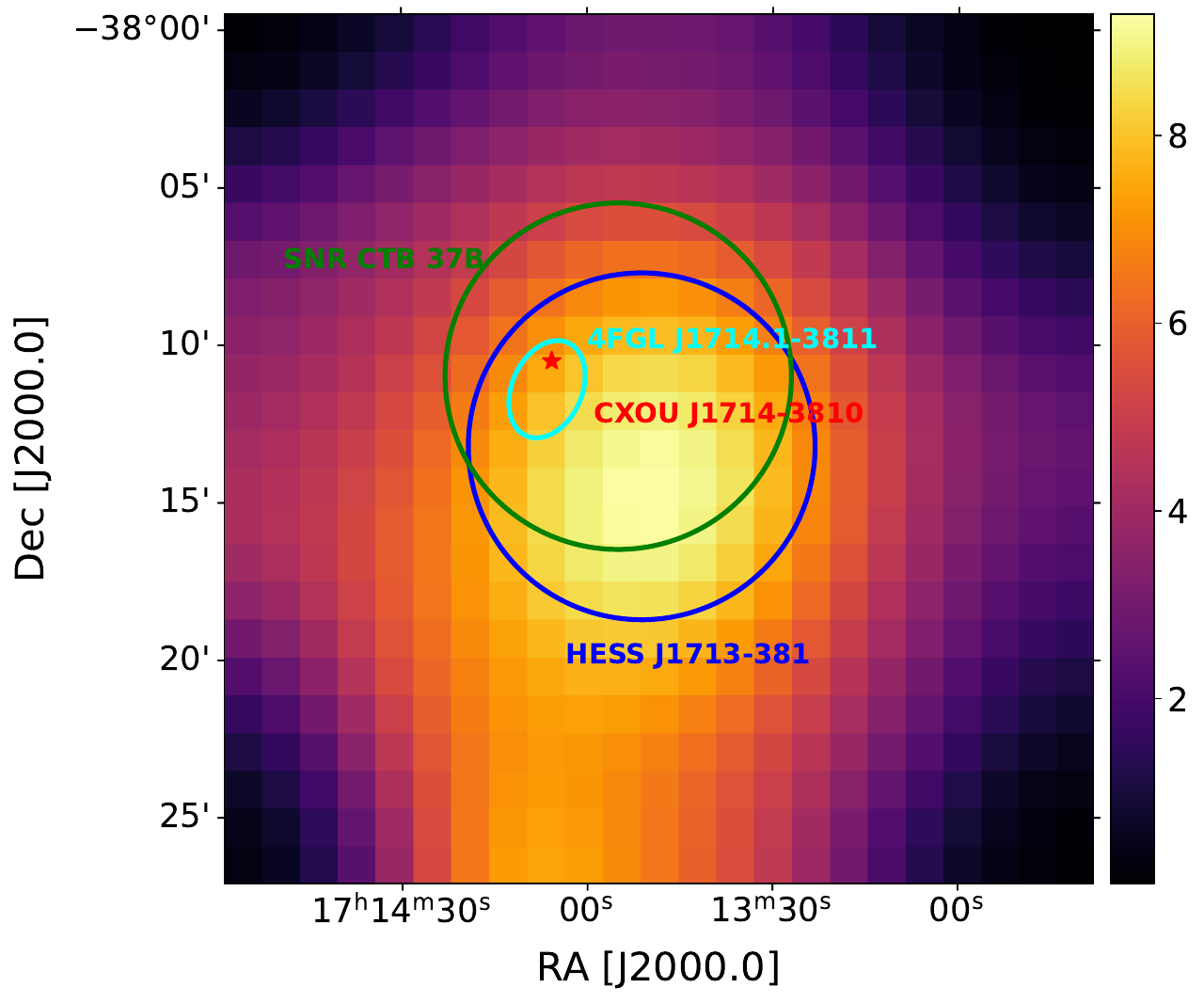}
    \caption{H.E.S.S. VHE $\gamma$-ray significance map of the region surrounding SNR CTB~37B / CXOU~J1714--3810 \citep{2018A&A...612A...1H}, smoothed with a Gaussian kernel of width $0.05^{\circ}$. The red star marks the position of the magnetar CXOU~J1714--3810 \citep{2014ApJS..212....6O}. The green circle indicates the radio extent of CTB~37B, with a radius of $5.1'$ \citep{2025JApA...46...14G, 2016ApJ...817...64X}. The cyan ellipse represents the GeV counterpart detected by \textit{Fermi}-LAT \citep[$68\%$ positional uncertainty;][]{2022ApJS..260...53A}, and the blue circle corresponds to the $1\sigma$ extension of the TeV source detected by H.E.S.S. \citep{2018A&A...612A...1H}.}
    \label{fig:signif_map}
\end{figure}

CXOU~J1714--3810 is a magnetar located within the SNR CTB~37B \citep[see Fig.~\ref{fig:signif_map} and][]{2008A&A...490..685A}. This magnetar exhibits a spin period of approximately $3.82$~s and a spin-down rate of $\dot{P} = (6.40 \pm 0.05) \times 10^{-11}$~s~s$^{-1}$ \citep{2010PASJ...62L..33S}, indicating a strong surface magnetic field of approximately $5 \times 10^{14}$~G. The characteristic age derived from these parameters suggests that CXOU~J1714--3810 is a relatively young magnetar, with an estimated age of about $950$~yr \citep{2010PASJ...62L..33S}.

Observations have revealed both thermal and non-thermal X-ray emission from the region around CXOU~J1714--3810. X-ray observations conducted with \textit{Chandra} and \textit{XMM–Newton} identified extended diffuse thermal emission coincident with the radio shell of CTB~37B \citep{2008A&A...486..829A, 2019MNRAS.487.5019B}. The thermal X-ray spectrum is well described by a non-equilibrium ionization (NEI) model with an ionization timescale of $\tau = (7^{+9}_{-3}) \times 10^{10}$~s~cm$^{-3}$, indicating that the remnant is in a transition phase between shock heating and equilibrium. The estimated temperature of $kT \approx 0.8 \pm 0.1$~keV suggests an SNR age of approximately $4900$~yr \citep{2008A&A...486..829A}, assuming a distance of $7$~kpc \citep{1975A&A....45..239C, 2008A&A...486..829A}.

The external ambient density surrounding the remnant is estimated to be $\sim 0.5$~cm$^{-3}$, although this remains uncertain due to the complexity of the SNR environment. The presence of asymmetries in the X-ray emission suggests interaction with an inhomogeneous medium, potentially modifying shock dynamics and cosmic-ray acceleration efficiency. The inferred density is consistent with independent estimates derived from thermal X-ray flux measurements, which support a scenario in which CTB~37B is expanding into a medium of moderate density \citep{2008A&A...486..829A}.

Recent X-ray investigations have focused on a region within CTB~37B exhibiting non-thermal emission to the south of the radio shell. The analysis favors a broken power-law model with photon indices of $1.23 \pm 0.23$ and $2.24 \pm 0.16$ below and above a break at $5.57 \pm 0.52$~keV, respectively, indicating a steepening spectrum. The inferred absorbing column density is consistent with that of the SNR, supporting a physical association. This spectral behavior may be explained by non-thermal bremsstrahlung emission from sub-relativistic electrons, although alternative interpretations, such as an unrelated pulsar wind nebula (PWN) or efficient shock acceleration, cannot be ruled out \citep{2024ApJ...977..163K}.

In the gamma-ray band, observations with the High Energy Stereoscopic System (\hess) and \textit{Fermi}-LAT have revealed sources spatially associated with CTB~37B (see Fig.~\ref{fig:signif_map}). At GeV energies, \textit{Fermi}-LAT detects the source 4FGL~J1714.1--3811 with a significance of $\sim 12.5\sigma$. Its position is coincident with the radio extent of CTB~37B, and it exhibits an intrinsic angular size of $\sim 1.6'$. The spectrum peaks at $\sim 10$~GeV, with an integral photon flux in the $1$--$100$~GeV range of $(1.55 \pm 0.23) \times 10^{-9}$~cm$^{-2}$~s$^{-1}$. At TeV energies, \hess\ detects the source HESS~J1713--381, also spatially coincident with the SNR. The TeV emission presents a larger intrinsic extension of $\sim 5.5' \pm 1.0'$ and is described by a power-law spectrum with photon index $\Gamma = 2.74 \pm 0.12$. The integral flux above $1$~TeV is $(0.52 \pm 0.07) \times 10^{-12}$~cm$^{-2}$~s$^{-1}$. The shell-like TeV morphology, together with the absence of diffuse non-thermal X-ray emission, favors a hadronic origin in which neutral pions from proton--proton collisions produce the observed gamma rays, rather than inverse-Compton scattering by relativistic electrons. This interpretation is consistent with the relatively soft TeV spectrum and with the ambient densities inferred for CTB~37B, although a leptonic contribution cannot be entirely excluded \citep{2008A&A...486..829A}. In this hadronic SNR picture, the gamma-ray emission arises from protons accelerated at the SNR forward shock, and the magnetar does not play a dynamical role in the acceleration process.

The proximity of CXOU~J1714–3810 to the peak of the \hess brightness contour, along with consistent absorption column densities between thermal and non-thermal X-ray emissions, supports a genuine association between the magnetar and CTB 37B. However, the lack of an extended X-ray nebula argues against a traditional PWN scenario. If CXOU~J1714–3810 is indeed associated with CTB 37B, its high transverse velocity \citep[$\sim 1000 (d/7~\textrm{kpc}) (t/5000~\textrm{years})^{-1}$~km/s; see][]{2008A&A...486..829A} would make it among the fastest-moving pulsars. Long-term X-ray monitoring over eight years has revealed variability in the emission from CXOU~J1714–3810, providing insights into the magnetar's activity and its interaction with the surrounding environment \citep[][]{2019ApJ...882..173G}.

\section{Source modeling and Results}  \label{sec:results_CXOU}

In this section, we investigate the region around the magnetar CXOU~J1714--3810 using both leptonic and lepto-hadronic models, applying MCMC techniques to fit radiative models via the \texttt{Naima} software \citep{naima}. The modeling process integrates multiwavelength observations, spanning radio frequencies to very-high-energy gamma rays. To characterize the populations of relativistic electrons and protons, we employ two particle distribution models: an exponential cutoff power-law (\textit{ecpl}) and a log-parabolic (\textit{lp}) model, which provide adequate phenomenological descriptions for the energy spectra of accelerated particles. The \textit{ecpl} model is given by:
\begin{equation}
\Phi(E) = \Phi_{0} \left( \frac{E}{E_{0}} \right)^{-\Gamma} \exp\left( -\frac{E}{E_{\rm cut}} \right),
\label{eq:ecpl}
\end{equation}
while the \textit{lp} model is expressed as:
\begin{equation}
\Phi(E) = \Phi_{0} \left( \frac{E}{E_{0}} \right)^{-\Gamma - \beta \log\left( \frac{E}{E_{0}} \right)},
\label{eq:lp}
\end{equation}
where $\Phi(E)$ denotes the flux at energy $E$, $\Phi_{0}$ is the flux normalization at the reference energy $E_{0}$, $\Gamma$ is the spectral index, $E_{\rm cut}$ is the energy cutoff, and $\beta$ describes the spectral curvature. In the lepto-hadronic modeling, the flux normalizations of the electron and proton populations are scaled relative to each other using the normalization ratio $K_{\rm ep} = \Phi_{\rm 0,e}/\Phi_{\rm 0,p}$ \citep{2012A&A...538A..81M} at a reference energy of $1$~TeV.

In both leptonic and lepto-hadronic scenarios, radio-to-X-ray emission is attributed to synchrotron radiation from relativistic electrons interacting with the local magnetic field. The gamma-ray emission, however, originates from IC scattering of background photon fields in the leptonic model and from neutral-pion ($\pi^0$) decay in the hadronic model. The radiation field includes contributions from the cosmic microwave background (CMB), an infrared component with a temperature of $T = 30$~K and an energy density of $u = 1$~eV~cm$^{-3}$ \citep{2006ApJ...648L..29P}, and an optical component with $T = 6000$~K and $u = 1$~eV~cm$^{-3}$ \citep{2006ApJ...648L..29P}. The ambient gas density, inferred from X-ray data, is assumed to be $0.5$~cm$^{-3}$ \citep{2008A&A...486..829A, 2009PASJ...61S.197N}.

\begin{table*}
\centering
\renewcommand{\arraystretch}{1.5} 
\caption{Spectral fitting parameters for the leptonic and lepto-hadronic models describing the multiwavelength emission from the CXOU~J1714--3810 region in the SNR and PWN scenarios. The columns list the model type, electron-to-proton ratio ($K_{\rm ep}$), electron and proton spectral indices ($\Gamma_{\rm e}$, $\Gamma_{\rm p}$), cutoff energies for electrons and protons ($E_{\rm e,cut}$, $E_{\rm p,cut}$), spectral curvature parameters ($\beta_{\rm e}$, $\beta_{\rm p}$) where applicable, total energy in relativistic electrons ($W_{\rm e}$) and protons ($W_{\rm p}$), magnetic field strength ($B$), and Bayesian Information Criterion (BIC) values for statistical model comparison. Lower BIC values indicate more statistically favored models.}
\label{tab:Spec_model_cxou}
\vspace{0.4cm}

\resizebox{\textwidth}{!}{%
\begin{tabular}{lccccccccccc}
\hline
\hline
MODEL & \multicolumn{1}{l}{$K_{\rm ep}$} & $\Gamma_e$ & $\Gamma_p$ & \begin{tabular}[c]{@{}c@{}}$E_{e, \mathrm{cut}}$ \\ (TeV)\end{tabular} & \begin{tabular}[c]{@{}c@{}}$E_{p, \mathrm{cut}}$\\ (TeV)\end{tabular} & $\beta_e$ & $\beta_p$ & \begin{tabular}[c]{@{}c@{}}$W_e$\\ ($10^{50}$ erg)\end{tabular} & \begin{tabular}[c]{@{}c@{}}$W_p$\\ ($10^{50}$ erg)\end{tabular} & \begin{tabular}[c]{@{}c@{}}$B$\\ ($\mu$G)\end{tabular} & BIC \\ 
\hline
\multicolumn{12}{c}{CXOU~J171405-381031 region in the SNR scenario} \\ 
\hline
Leptonic-{\it ecpl} & --- & $2.18 \pm 0.04$ & --- & $3.17 \pm 0.56$ & --- & --- & --- & 0.213 & --- & $27.6 \pm 2.9$ & 47.92 \\
Leptonic-{\it lp} & --- & $2.88 \pm 0.04$ & --- & --- & --- & $0.085 \pm 0.008$ & --- & 0.409 & --- & $16.7 \pm 1.7$ & 50.76 \\
Lepto-hadronic-{\it ecpl} & $10^{-2}$ & $2.20 \pm 0.04$ & $2.04 \pm 0.32$ & $3.06 \pm 1.31$ & $122.8 \pm 90.2$ & --- & --- & 0.178 & 12.21 & $29.6 \pm 3.5$ & 49.91 \\
Lepto-hadronic-{\it lp} & $10^{-2}$ & $3.17 \pm 0.17$ & $0.12 \pm 0.72$ & --- & --- & $0.117 \pm 0.020$ & $0.673 \pm 0.302$ & 0.539 & 5.50 & $13.6 \pm 2.3$ & 54.42 \\
\hline
\multicolumn{12}{c}{CXOU~J171405-381031 region in the PWN scenario} \\ 
\hline
Leptonic-{\it ecpl} & --- & $2.28 \pm 0.23$ & --- & $3.75 \pm 2.03$ & --- & --- & --- & 0.296 & --- & $5.18^{+12.6}_{-3.80}$ & 19.33 \\
Leptonic-{\it lp} & --- & $2.69 \pm 0.14$ & --- & --- & --- & $0.29 \pm 0.10$ & --- & 0.062 & --- & $11.3^{+12.2}_{-7.92}$ & 18.47 \\
\hline
\hline
\end{tabular}%
}
\end{table*}

\subsection{CXOU~J1714–3810 region in the SNR scenario} \label{sec:snr_scenario}

We first investigated whether the high- and very-high-energy gamma-ray emission observed in the region around the magnetar CXOU~J1714--3810 originates from cosmic rays accelerated in the supernova remnant CTB~37B. SNRs are well established as efficient accelerators of Galactic cosmic rays via diffusive shock acceleration (DSA) \citep[see e.g.][]{1978ApJ...221L..29B, 1978MNRAS.182..147B, 1983RPPh...46..973D, 2008ARA&A..46...89R, 2013Sci...339..807A}. For our modeling, we adopted a distance to CTB~37B of $13.2$~kpc \citep{2012MNRAS.421.2593T}, corresponding to a physical radius of $20$~pc and an angular size of $5.1$~arcmin at this distance \citep{2016ApJ...817...64X}. 

Gamma-ray data are fitted using observations from \textit{Fermi}-LAT and \hess\ \citep{2017ApJS..232...18A, 2022ApJS..260...53A, 2018A&A...612A...1H}, while synchrotron emission is modeled based on radio observations of the shell-type SNR CTB~37B \citep{1991ApJ...374..212K}. Since no significant non-thermal X-ray emission has been detected in the region coincident with the SNR shell, the observed thermal X-ray flux in the $2$--$10$~keV range \citep[see][]{2008A&A...486..829A, 2019MNRAS.487.5019B} is adopted as an upper limit on any potential non-thermal component.

\begin{figure*} 
    \centering
    \begin{subfigure}{0.49\textwidth}
        \includegraphics[width=\linewidth]{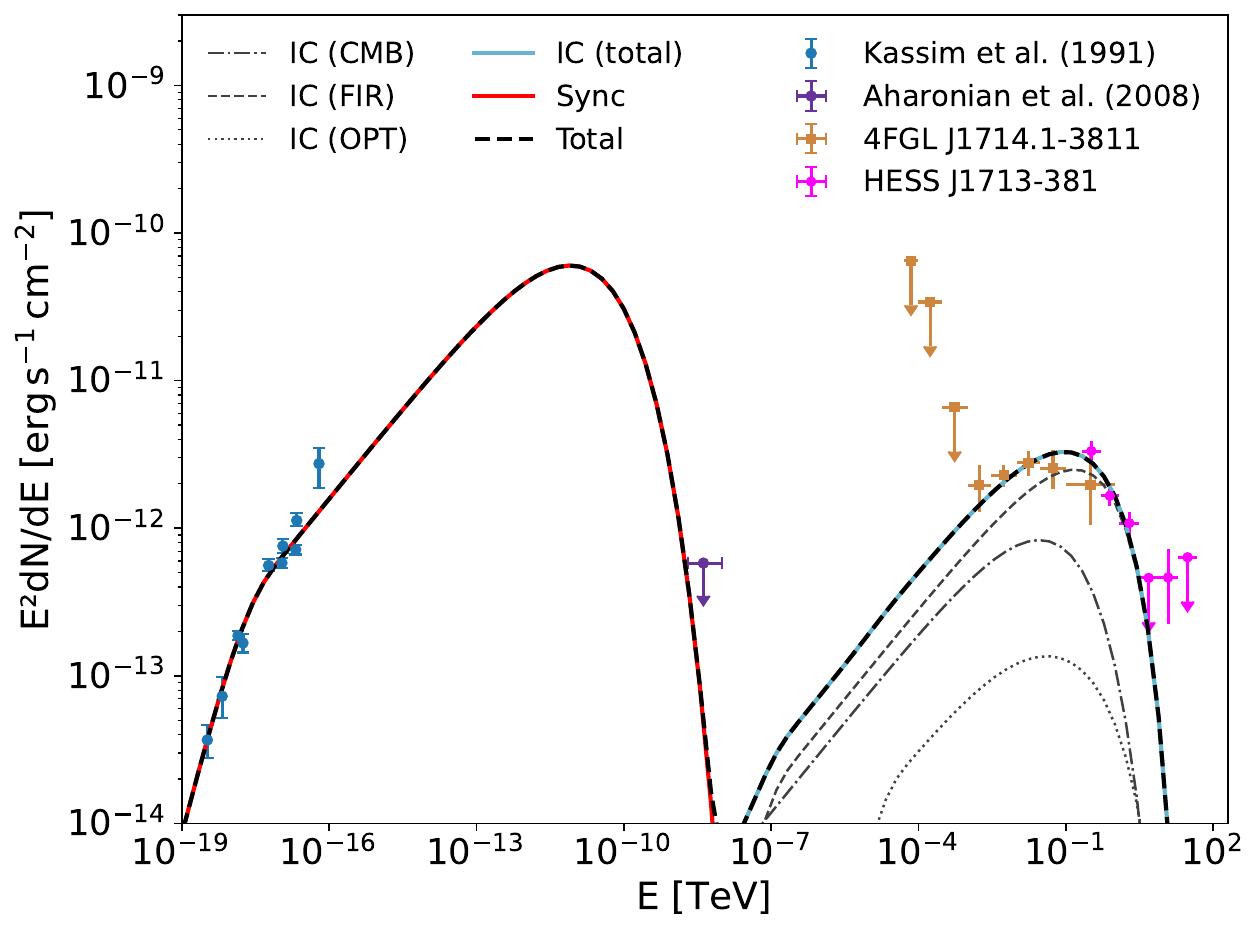}
        \caption{Leptonic-{\it ecpl} model}
    \end{subfigure}
    \hfill
    \begin{subfigure}{0.49\textwidth}
        \includegraphics[width=\linewidth]{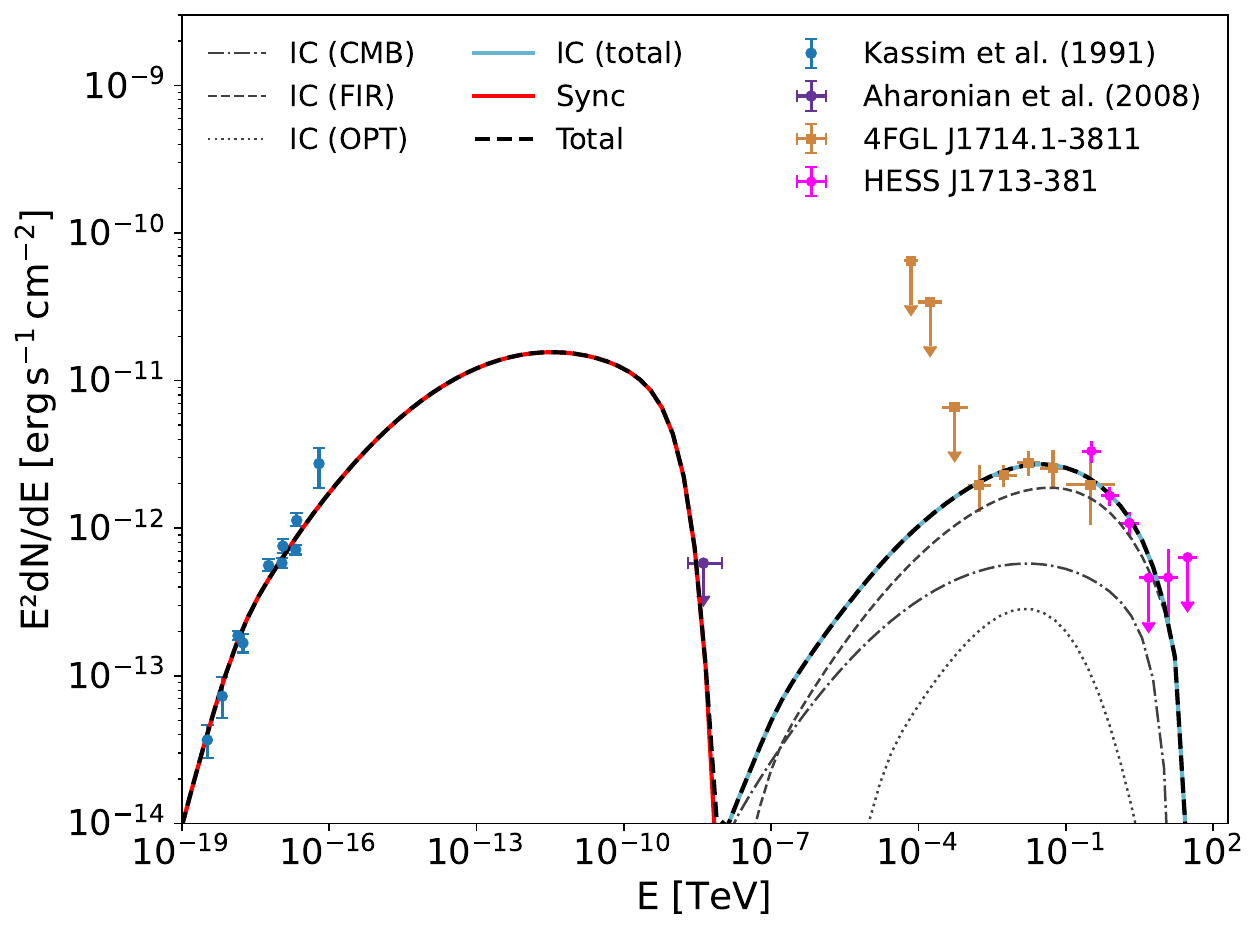}
        \caption{Leptonic-{\it lp} model}
    \end{subfigure}
    \caption{Multi-band SED of the region around the magnetar CXOU~J1714-3810 in the SNR model for leptonic models. Observational data across radio, X-ray, GeV, and TeV bands are taken from \citet{1991ApJ...374..212K} (radio), \citet{2008A&A...486..829A, 2019MNRAS.487.5019B} (X-ray), \citet{2017ApJS..232...18A, 2022ApJS..260...53A} (GeV), and \citet{2018A&A...612A...1H} (TeV). The right panel presents the SED modeled with an electron distribution following a log-parabolic ({\it lp}) function, while the left panel shows the SED using an exponential cutoff power-law ({\it ecpl}) model for the electron spectrum.}
    \label{fig:CXOU_IC}
\end{figure*}

\begin{figure*} 
    \centering
    \begin{subfigure}{0.49\textwidth}
        \includegraphics[width=\linewidth]{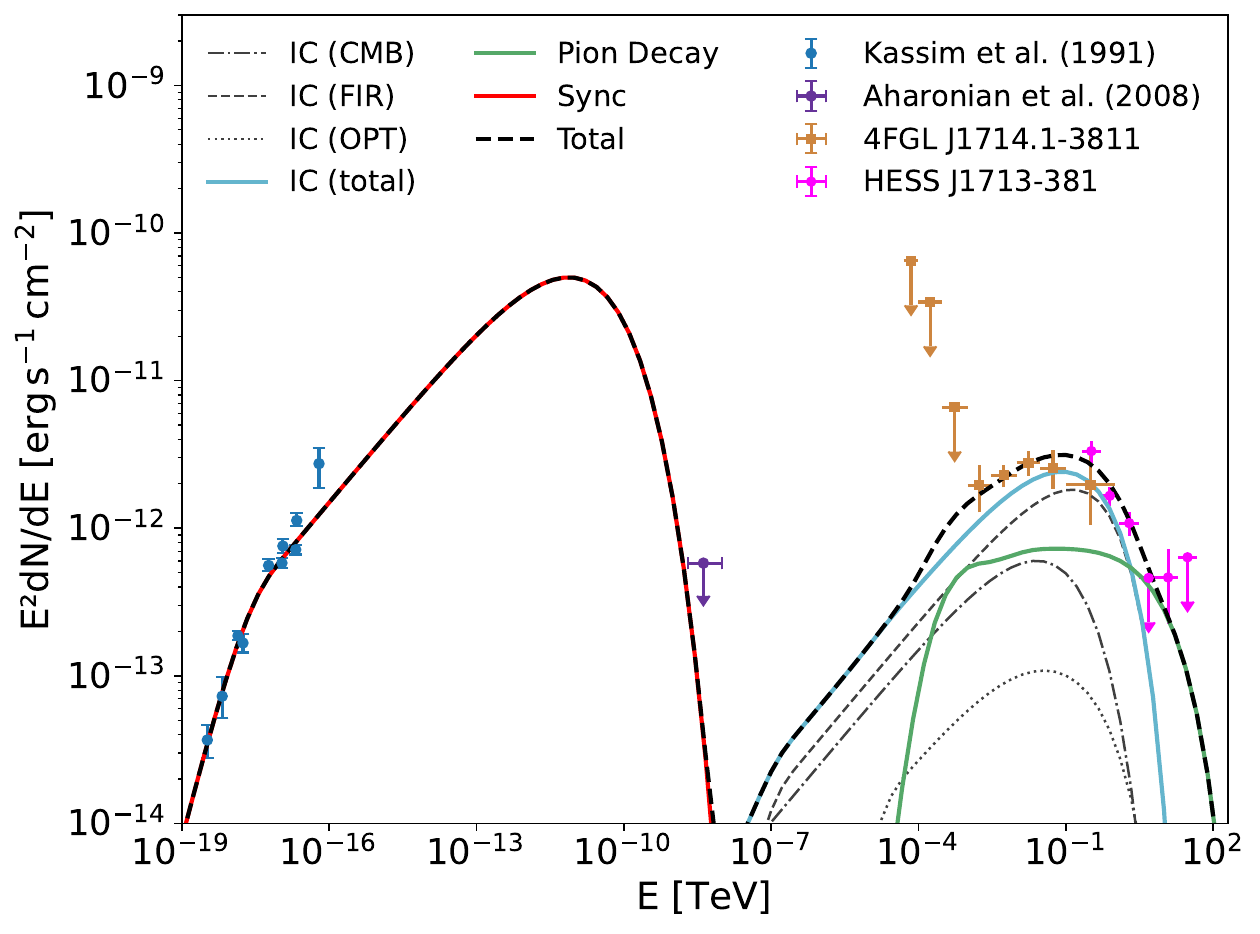}
        \caption{Lepto-hadronic-{\it ecpl}}
    \end{subfigure}
    \hfill
    \begin{subfigure}{0.49\textwidth}
        \includegraphics[width=\linewidth]{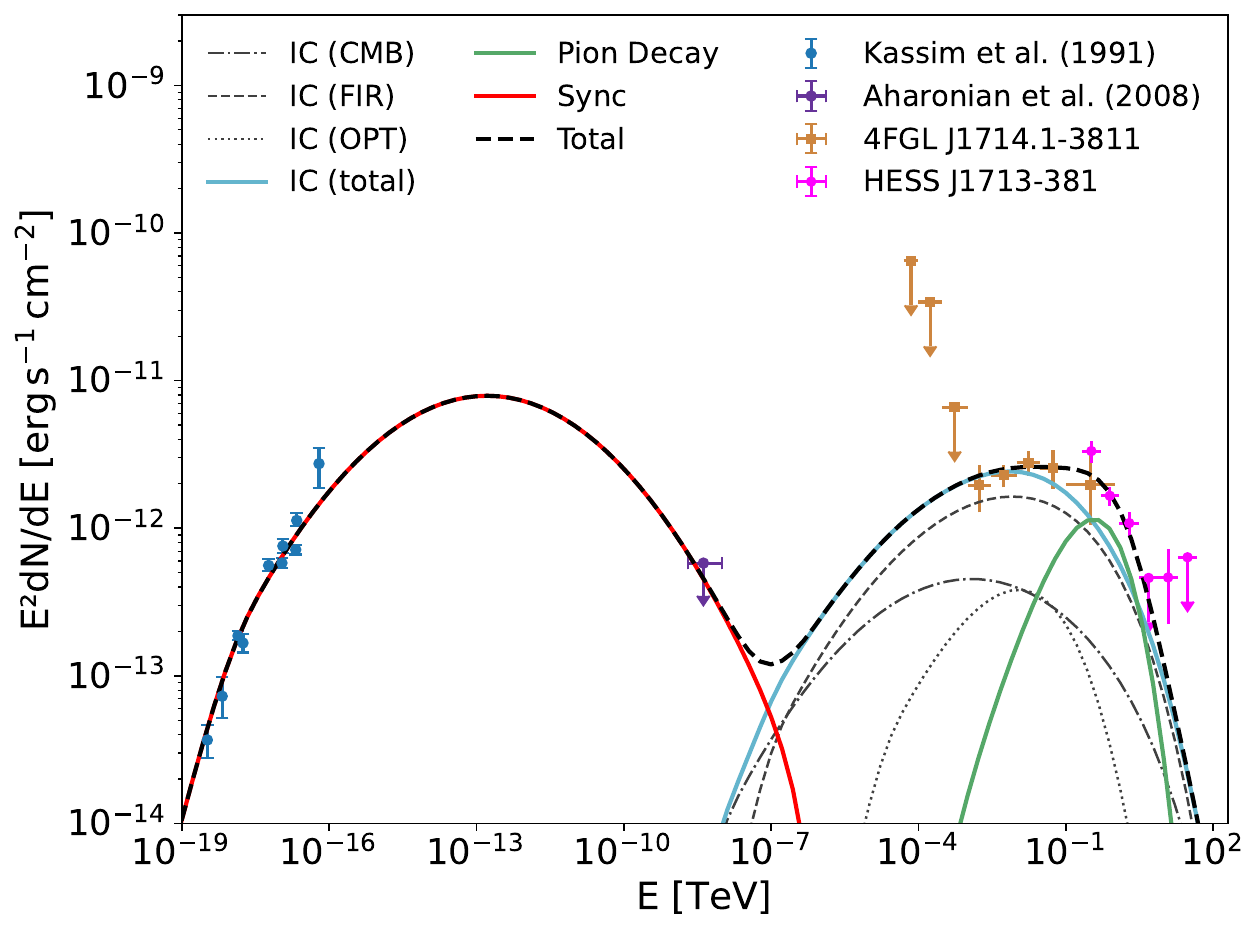}
        \caption{Lepto-hadronic-{\it lp}}
    \end{subfigure}
    \caption{Multi-band SED of the magnetar CXOU~J1714-3810 region modeled within the SNR lepto-hadronic scenario. The relativistic electron and proton populations are described by an exponential cutoff power-law distribution ({\it ecpl}; left panel) and a log-parabolic distribution ({\it lp}; right panel). In both cases, a fixed electron-to-proton ratio of $K_{\rm ep} = 10^{-2}$ is assumed. }
    \label{fig:CXOU_PD_IC_ecpl}
\end{figure*}

Table~\ref{tab:Spec_model_cxou} presents the spectral fitting parameters for four configurations used to model the multi-wavelength emission from the CXOU~J1714-3810 region within the SNR scenario: two purely leptonic and two lepto-hadronic scenarios. The leptonic models assume electron distributions described by either an {\it ecpl} or a {\it lp} function. Similarly, the lepto-hadronic models incorporate these same spectral shapes for both the electron and proton populations, assuming a fixed electron-to-proton ratio of $K_{\rm ep} = 10^{-2}$. Although alternative values of $K_{\rm ep}$ were explored, a ratio of $10^{-2}$ yielded the optimal fit to the observational data.

\subsubsection{Leptonic Model} 

Figure~\ref{fig:CXOU_IC} illustrates the spectral fit results for the leptonic model, highlighting the SED contributions from each radiative process. The high-energy photon emission is primarily driven by the interaction of relativistic electrons with the infrared photon field, with a smaller contribution from CMB photons. In the leptonic-\textit{ecpl} scenario (Fig.~\ref{fig:CXOU_IC}-a), the non-thermal X-ray component is weak and exhibits a rapidly declining spectrum. In contrast, in the leptonic-\textit{lp} scenario (Fig.~\ref{fig:CXOU_IC}-b), the electron spectrum is constrained to a maximum energy of $\sim 30$~TeV, since extending the distribution to higher energies would produce detectable synchrotron X-ray emission, which is not observed. Such a sharp cutoff in the electron spectrum is atypical and would require an additional physical mechanism to suppress particles above this energy. Given the inferred spectral index and curvature of the \textit{lp} distribution, electrons with energies exceeding 30~TeV would be expected to be present, albeit in smaller numbers, unless an efficient high-energy loss or escape process is at work.

The models can be evaluated using the Bayesian Information Criterion (BIC; see Table~\ref{tab:Spec_model_cxou}) to identify the statistically preferred framework. Between the two leptonic scenarios, the {\it ecpl} distribution provides a marginally better statistical fit; its lower BIC value yields $\Delta{\rm BIC} \approx 2.84$ relative to the {\it lp} model, corresponding to a relative probability\footnote{The relative probability that a model 1 is statistically preferred over a model 2 is given by $P = 1/(1+\exp{(-\Delta{\rm BIC}/2)})$, where the difference $\Delta{\rm BIC} = {\rm BIC}_2 - {\rm BIC}_1$ \cite[see e.g.,][]{2007MNRAS.377L..74L, 2017ApJ...834..153Z}} of $P \approx 0.80$. Following the strength-of-evidence scale proposed by \citet{2008ConPh..49...71T}\footnote{Following the strength-of-evidence scale proposed by \citet{2008ConPh..49...71T}, differences in BIC values are interpreted as $\Delta{\rm BIC} = 2$–$5$ indicating weak evidence, $5$–$10$ indicating moderate evidence, and $>10$ indicating strong evidence in favor of the model with the lowest BIC value \citep[see also][]{kass1995bayes}.}, this result constitutes weak statistical evidence to distinguish between the two models, rendering the leptonic-{\it ecpl} scenario only slightly preferred. Nevertheless, the {\it lp} model introduces an atypical cutoff in the maximum energy of the electron distribution to accommodate the X-ray upper limit constraint, favoring the leptonic-{\it ecpl} framework as the more robust description of the data.


\subsubsection{Lepto-hadronic Model} 

Figure~\ref{fig:CXOU_PD_IC_ecpl} presents the spectral fitting results for the lepto-hadronic scenarios, emphasizing the relative contributions of the different radiative processes to the overall emission. In the lepto-hadronic configuration assuming an {\it ecpl} distribution, the gamma-ray emission from pion decay is comparable to that from the IC process, with pion decay dominating within a narrow spectral window at energies above $10$~TeV (Fig.~\ref{fig:CXOU_PD_IC_ecpl}-a). Conversely, for the {\it lp} distribution (Fig.~\ref{fig:CXOU_PD_IC_ecpl}b), IC scattering remains the primary emission mechanism across nearly the entire gamma-ray spectrum, though pion decay contributes significantly within the $0.1-10$~TeV range. Regarding the synchrotron component, both models yield a non-thermal X-ray flux comparable to the observed thermal emission; however, the synchrotron spectrum exhibits a much sharper high-energy cutoff for the {\it ecpl} model than for the broader {\it lp} distribution.

The total energy content of the relativistic particles in each scenario, calculated for energies above $1$~GeV, is presented in Table~\ref{tab:Spec_model_cxou}. The total energy of relativistic electrons ($W_e$) is on the order of $10^{49}$~erg. This estimate of the particles' energy reserve is roughly an order of magnitude higher than what is typically observed in other gamma-ray-emitting SNRs, such as RX~J1713–3946 \citep{2011ApJ...735..120Y}, RX~J0852–4622 \citep{2011ApJ...740L..51T}, and HESS~J1731–347 \citep{2014A&A...567A..23Y}. For relativistic protons, the total energy ($W_p$) exceeds $10^{50}$~erg, a remarkable value considering that the typical kinetic energy released by a core-collapse supernova is around $10^{51}$~erg. Such elevated energy levels could indicate an exceptionally massive progenitor star for CTB~37B or a magnetar-powered supernova event \citep[see Sect.~\ref{sec:discussions_CXOU} and][]{2015Natur.523..189G, 2016ApJ...817...64X, 2020NatAs...4..893N}.

For the lepto-hadronic scenarios, the model employing the {\it ecpl} function is statistically favored over the {\it lp} model, with $\Delta{\rm BIC} \approx 4.51$, corresponding to a relative probability of $P \approx 0.90$. According to the adopted BIC interpretation scale, this difference represents weak-to-moderate evidence in favor of the {\it ecpl} model. This result is consistent with the spectral fits shown in Fig.~\ref{fig:CXOU_PD_IC_ecpl}, where this scenario effectively reproduces the observed gamma-ray emission, despite the relatively high energy budget required for relativistic particles compared to the typical values inferred for other gamma-ray–emitting SNRs.

Beyond the statistical preference indicated by the BIC analysis, the physical implications of the best-fit parameters strongly disfavor the lepto-hadronic-{\it lp} model. This scenario requires an extremely hard proton spectral index ($\Gamma_p \approx 0.12$), which is difficult to reconcile with standard particle acceleration theories in supernova remnants. In the framework of DSA for strong shocks, the predicted injection index is $\approx 2.0$ \citep{1978MNRAS.182..147B, 1978ApJ...221L..29B}. Nonlinear DSA effects can produce concave spectra that harden at the highest energies, yet the asymptotic hadronic index typically does not fall below $\sim 1.5$ \citep[see e.g.,][]{1999ApJ...526..385B, 2001RPPh...64..429M, 2012JCAP...07..038C}. An index as extreme as $0.12$ implies a nearly flat particle distribution, one whose energy content diverges non-physically toward the cutoff region. The unphysical nature of this spectral hardening therefore provides a compelling theoretical justification to discard the {\it lp} model, reinforcing the choice of the {\it ecpl} scenario.

\subsection{CXOU~J1714–3810 region in the MWN scenario} 
\label{sec:PWN_model_CXOU}

Previous studies have suggested that the GeV–TeV gamma-ray emission in the CXOU~J1714–3810 region originates from particles accelerated at the shock front of the SNR, either through pion decay or via IC scattering off background photon fields \citep{2008A&A...486..829A, 2009PASJ...61S.197N, 2016ApJ...817...64X}. Our results based on the SNR scenario also support this interpretation, demonstrating that the observed high-energy emission can be effectively modeled within a leptonic or a lepto-hadronic framework.

Nevertheless, a potential contribution from the magnetar cannot be entirely ruled out. One plausible scenario is that the observed high-energy flux includes an IC component produced by an undetected MWN confined within the SNR shell. Unlike ordinary pulsars, the capacity of magnetars to accelerate particles to TeV energies remains uncertain. Magnetar models involving strong currents along closed and twisted magnetic field lines predict voltage drops on the order of $\approx 10^{9}$~V \citep{2007ApJ...657..967B}, which may be insufficient to energize particles to the TeV scale. On the other hand, it is plausible that the standard pulsar mechanism could operate along open magnetic field lines of magnetars. In this context, magnetars might produce particle-dominated winds capable of forming shocked MWNe, thereby accelerating a fraction of the particles to TeV energies \citep{2010ApJ...725.1384H}.

\begin{figure*} 
    \centering
    \begin{subfigure}{0.49\textwidth}
        \includegraphics[width=\linewidth]{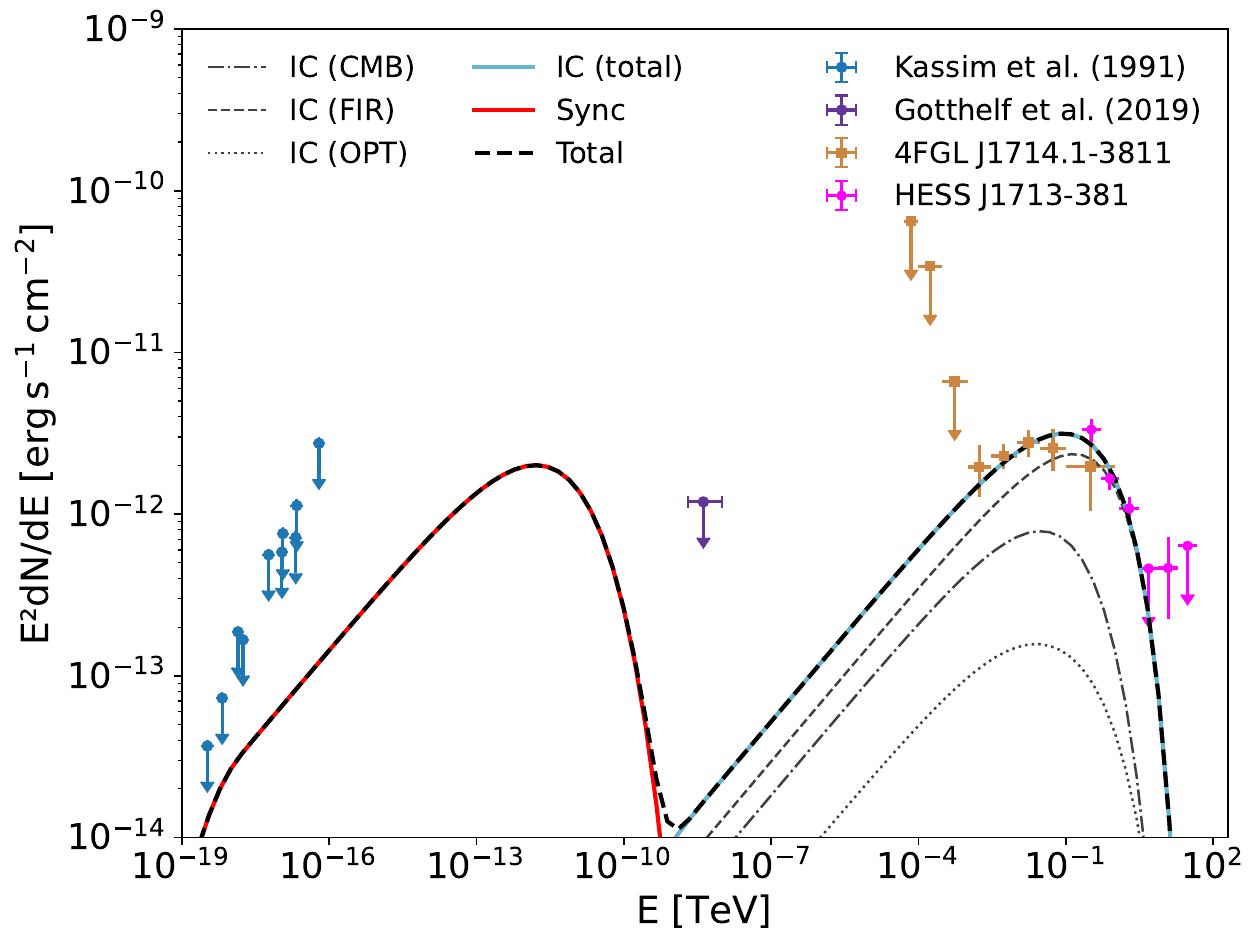}
        \caption{Leptonic-{\it ecpl} model}
    \end{subfigure}
    \hfill
    \begin{subfigure}{0.49\textwidth}
        \includegraphics[width=\linewidth]{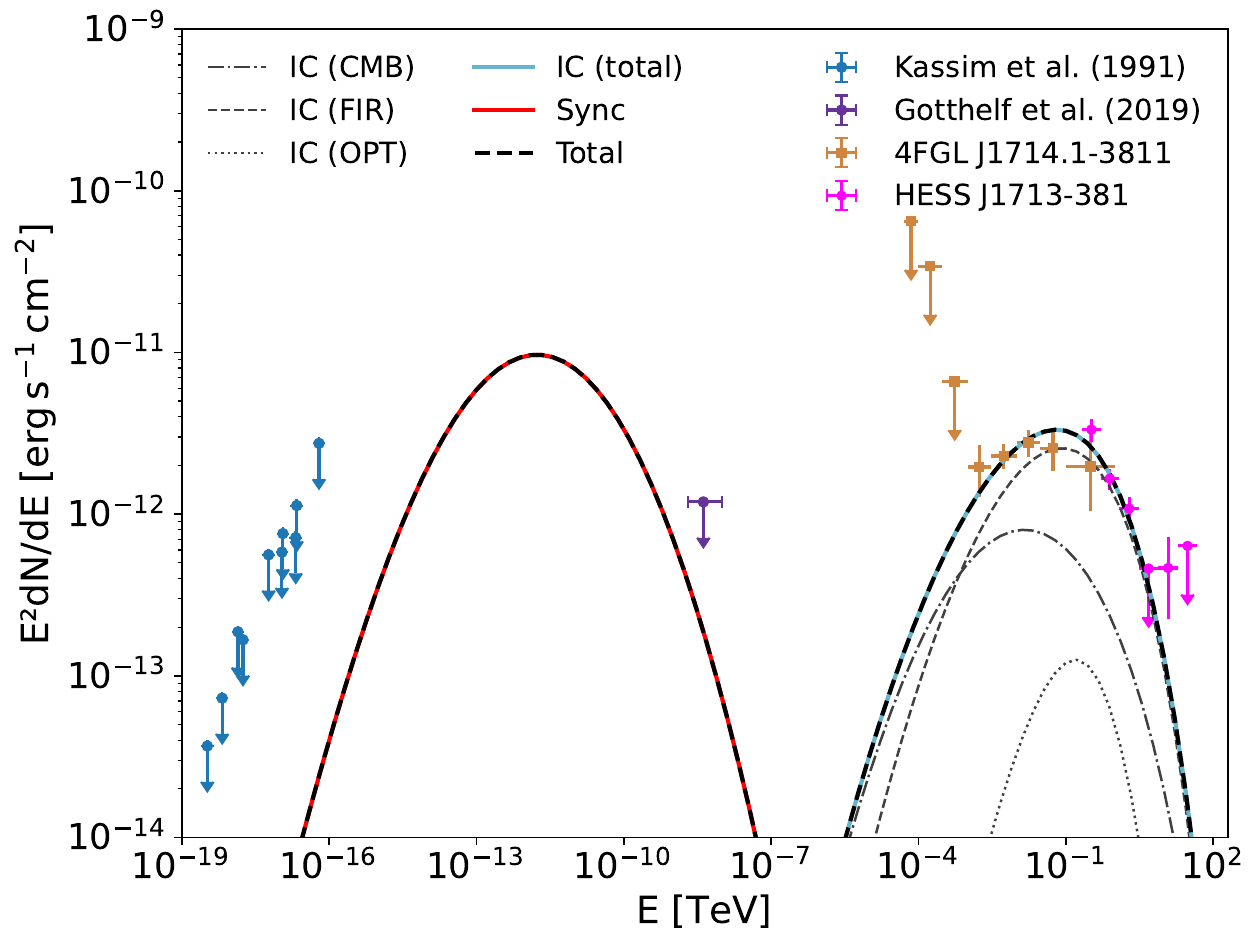}
        \caption{Leptonic-{\it lp} model}
    \end{subfigure}
    \caption{Multi-wavelength SED of the magnetar CXOU~J1714-3810 region within the MWN model. The left panel shows the SED modeled with an electron spectrum following an {\it ecpl}, while the right panel uses a {\it lp} distribution. Observational data are from \citet{1991ApJ...374..212K} (radio), \citet{2019ApJ...882..173G} (X-ray), \citet{2017ApJS..232...18A, 2022ApJS..260...53A} (GeV), and \citet{2018A&A...612A...1H} (TeV).}
    \label{fig:CXOU_IC_PWN}
\end{figure*}

Interestingly, a wind nebula producing extended X-ray emission was recently detected around the magnetar Swift~J1834–0846 \citep{2016ApJ...824..138Y}, strengthening the hypothesis that magnetars can form MWNe. As demonstrated by \cite{2017ApJ...835...54T}, this nebula can be interpreted as a rotationally powered MWN, similar to those seen around conventional pulsars, provided it is currently being compressed by the surrounding medium.

Motivated by these findings, we investigate whether the high- and very-high-energy gamma-ray emission around CXOU~J1714–3810 can be attributed to a leptonic scenario in which electrons are accelerated at the MWN termination shock, assuming a source distance of 13.2~kpc.

Since there is no direct X-ray evidence of a MWN associated with CXOU~J1714–3810, we adopt the X-ray flux measurements obtained by {\it XMM-Newton} and {\it NuSTAR} for the magnetar \citep{2019ApJ...882..173G} as an upper limit on the possible MWN emission. Likewise, the radio data \citep{1991ApJ...374..212K} associated with the SNR are also treated here as upper limits, since any radio flux exceeding that of the SNR shock would likely have been detected; however, observations show no radio emission from regions other than the SNR shock. Therefore, these upper limit flux points are used to constrain the synchrotron component of the relativistic electron population in our simulations.

Table~\ref{tab:Spec_model_cxou} summarizes the spectral fitting parameters for two leptonic models used to describe the multiwavelength emission from the CXOU~J1714–3810 region within the MWN model. These models employ electron energy distributions characterized by either an {\it ecpl} or a {\it lp} function.

Figure~\ref{fig:CXOU_IC_PWN} displays the corresponding SEDs for these two models, illustrating the individual contributions of each radiative process. The high-energy gamma-ray emission is primarily produced through IC scattering of infrared background photons by relativistic electrons, with a smaller contribution from scattering off CMB photons. The fitted curves provide a good representation of the high-energy observational data. However, the magnetic field strength inferred from the synchrotron component is poorly constrained due to the lack of sufficient observational data at lower energies. This limitation is reflected in Table~\ref{tab:Spec_model_cxou}, where the uncertainties associated with the magnetic field estimates are large relative to the central values.

Given the young age of the system, its particle population is expected to be dominated by freshly accelerated particles rather than a highly cooled, evolved distribution. While a one-zone model technically integrates both injected and cooled particles, fresh injection should dominate at such an early epoch, so the electron spectral index inferred from our modeling serves as a proxy for the intrinsic injection spectrum. The value obtained from our {\it ecpl} fit ($\Gamma_e = 2.28$) aligns well with the injection parameters derived from detailed multi-zone modeling of young PWNe. For instance, \citet{2011ApJ...742...62V} adopted an injection index of $2.2$ for their evolutionary models, while the multi-zone analysis by \citet{2023ApJ...945...66P} yielded a comparable particle injection index of $2.27$ for J1418$-$6058. This agreement indicates that the non-thermal emission in our scenario is driven by a recently injected, uncooled particle distribution.

While the BIC (Table~\ref{tab:Spec_model_cxou}) nominally favors the leptonic model with a {\it lp} electron distribution over the {\it ecpl} model, the statistical preference is inconclusive ($\Delta{\rm BIC} \approx 0.85$). The relative probability of $P \sim 0.60$ signifies only a marginal improvement in the fit.

\section{Discussion}  \label{sec:discussions_CXOU}

Our analysis of the region surrounding CXOU~J1714-3810 within the SNR scenario suggests that both leptonic and lepto-hadronic models can adequately reproduce the observed broadband spectrum. However, based on the BIC, the leptonic-{\it ecpl} model emerges as the preferred purely leptonic scenario, while the lepto-hadronic-{\it ecpl} model is statistically favored within the hybrid framework.

\citet{2016ApJ...817...64X} examined this region using purely leptonic and purely hadronic scenarios, assuming an ambient gas density of $0.5\text{ cm}^{-3}$. They found that a leptonic scenario required a total electron energy above $1$~GeV approximately ten times higher than that observed in other SNRs that emit gamma rays. In the hadronic case, they estimated that the total energy of the relativistic protons exceeded $10^{51}$~erg. Assuming identical spectral indices for the electron and proton populations, \citet{2017ApJ...834..153Z} proposed a hybrid lepto-hadronic scenario in which the GeV emission arises from comparable contributions of hadronic interactions and IC scattering, while the TeV emission is predominantly dominated by hadronic processes. Our analysis broadens the range of interpretations for the region around the magnetar/SNR system by incorporating the latest high-energy observations and exploring multiple configurations within the SNR framework to statistically determine the best fit to the observational data. Furthermore, unlike \citet{2017ApJ...834..153Z}, our lepto-hadronic analysis allows the spectral indices of the particle distributions to vary independently.

\subsection{SNR scenario: leptonic model}

Previous studies \citep{2008A&A...486..829A} have argued that the TeV emission from CTB 37B region is of hadronic origin, specifically from proton--proton interactions occurring within the SNR shell. According to their interpretation, the lack of detected non-thermal X-ray emission disfavors a leptonic scenario unless either an unusually weak magnetic field ($B \sim 1~\mu$G) is assumed or the electron spectrum exhibits a sharp cutoff at a maximum energy of $\sim 40$~TeV. A similar requirement emerges in our leptonic-\textit{lp} model (see Sect.~\ref{sec:snr_scenario}), where the electron distribution must be truncated at $E_{\rm e,max} \sim 30$~TeV to remain consistent with the X-ray upper limits.

Our results further indicates that a purely leptonic model described by an \textit{ecpl} electron distribution can also account for the observed high-energy emission, although it requires a relatively weak magnetic field ($B \sim 30~\mu$G) compared to those typically inferred for young SNRs, which often exhibit fields of $B \sim 100$--$500~\mu$G (see discussion below). If, instead, a magnetic field of $B \sim 100~\mu$G with a spectral index of $\sim 2$ is assumed, the fit constrains the maximum electron energy to $E_{\rm e,max} \sim 7$~TeV to avoid producing detectable synchrotron X-ray emission. This implies a very sharp cutoff in the electron spectrum, given that the exponential cutoff already becomes effective at $E_{\rm e,cut} \sim 3$~TeV. Such behaviour in a high-magnetic-field environment is consistent with the results reported by \citet{2008A&A...486..829A}. The quantitative differences in the inferred cutoff energies likely arise from the inclusion of the most recent \hess observations and the incorporation of \textit{Fermi}-LAT data in our modelling, which provide additional constraints at GeV energies.

Notably, although the purely leptonic best-fit model underestimates the highest-energy flux point measured by \hess, it nevertheless provides a satisfactory overall description of the observed spectrum, yielding an electron cutoff energy of $E_{\rm e,cut} \sim 3$~TeV. This value is constrained by the upper limit on the synchrotron X-ray emission, as non-thermal X-rays would be expected from electrons with energies around $10$~TeV. The resulting spectral break at a few TeV produces a gamma-ray peak near $100$~GeV and a synchrotron peak in the ultraviolet range. However, this cutoff energy is lower than those typically observed in young SNRs, where values often reach tens of TeV. One possible explanation is that higher-energy electrons cool more rapidly if they were accelerated during the early stages of SNR evolution. Electrons at $\sim 50$~TeV in a $28~\mu$G magnetic field have a synchrotron cooling time significantly shorter than the estimated age of CTB~37B \citep[$\sim 5000$~yr;][]{2008A&A...486..829A}. In contrast, electrons around $3$~TeV have a cooling time of $t_{\rm sync} \sim 5300$~yr, consistent with the estimated age of CTB~37B, supporting the prediction that higher-energy electrons have already lost most of their energy via synchrotron and IC processes.

The leptonic model parameters found in this study resemble those of \citet{2016ApJ...817...64X}, with two notable differences. First, their model adopts a harder electron spectrum ($\Gamma_{\rm e} = 1.65$) compared to the value inferred here ($\Gamma_{\rm e} = 2.18$). Second, they infer a significantly stronger magnetic field ($B \simeq 100~\mu$G), whereas our best-fit model favors a lower field strength of $B \approx 28~\mu$G. In their scenario, the higher magnetic field is achieved at the expense of a harder electron spectral index. As noted by \citet{2017ApJ...834..153Z}, such hard spectra are difficult to reconcile with Galactic cosmic-ray compression, and nonlinear DSA generally does not produce a single power-law particle distribution with such steep hardening \citep[see also][]{2011JCAP...05..026C, 2012JCAP...07..038C, 2025MNRAS.544L.160S}. Our electron distribution resolves this spectral index discrepancy, yielding a value consistent with observational expectations, albeit at the expense of a weaker magnetic field.

It is also worth discussing the magnetic field strength inferred in our model within the broader context of SNR observations. From a general perspective, a field of order $B \sim 30~\mu$G is consistent with expectations for a more evolved SNR or for the compressed magnetic field, where the ambient interstellar magnetic field has been compressed by the passage of a strong shock \citep{2005JPhG...31R..95H, 2020pesr.book.....V}. However, this value becomes unusually low when compared with the magnetic fields inferred for the forward shocks of young SNRs such as RX J1713.7-3946 and Vela Junior \citep[see e.g.,][]{2007Natur.449..576U, 2005ApJ...632..294B, 2012ApJ...744...71I}. In those systems, the prominent non-thermal synchrotron X-ray emission observed along their thin rims provides clear evidence of substantial magnetic-field amplification. Constraints derived from the narrowness of the X-ray filaments often imply considerably larger magnetic fields, typically in the range of $100- 500~\mu$G, which are required to reproduce the observed X-ray morphology.

In contrast, CTB~37B does not exhibit strong non-thermal X-ray emission. The absence of such emission removes the observational requirement for an extremely amplified magnetic field, allowing the low-$B$ leptonic model to provide a consistent fit to the multiwavelength SED. Thus, while a field of $B \sim 30~\mu$G would be unusually low for a young SNR showing clear signs of efficient magnetic-field amplification, it remains fully compatible with the observational constraints on CTB~37B, which lacks prominent non-thermal X-ray morphological features. Similar magnetic field strengths have been inferred in other SNRs of comparable age, such as Puppis~A and HB9, further supporting the plausibility of this scenario \citep{2012ApJ...746...79A, 2013A&A...555A...9D, 2025A&A...697A.131X}.

\subsection{SNR scenario: lepto-hadronic model}

For the lepto-hadronic model, our best-fit results indicate that gamma-ray emission from GeV to a few TeV originates from a combination of IC scattering and pion decay, with IC slightly dominating at GeV energies. Above $10$~TeV, IC scattering becomes subdominant, and pion decay accounts for the highest-energy photons. This suggests that hadronic interactions are primarily responsible for the high-energy end of the spectrum.

The electron energy distribution in our model has a spectral index $\Gamma_e \sim 2.2$ and a cutoff energy of $E_{\rm e,cut} \sim 3$~TeV, leading to a gamma-ray peak near $100$~GeV. Meanwhile, the proton spectrum follows a spectral index $\Gamma_p \sim 2.0$ with a significantly higher cutoff at $E_{\rm p,cut} \sim 122$~TeV, implying that some protons are accelerated to hundreds of TeV to match the observed highest-energy gamma rays. Assuming that SNR~CTB~37B has a magnetic field strength of $B \sim 30~\mu$G and a physical radius of $\sim 20$~pc, it is capable of accelerating protons to these required high energies. These high cutoff values and spectral indices strongly support a scenario where CTB~37B serves as the primary gamma-ray source in the CXOU~J1714-3810 region, with particle acceleration driven by DSA \citep[see e.g.,][]{1978MNRAS.182..147B, 2007MPLA...22.1533H, 2011JCAP...05..026C, 2012JCAP...07..038C}. This model is further supported by the synchrotron cooling time ($t_{\rm sync} \sim 5200\text{ yr}$), which is compatible with the CTB~37B's estimated age.

We also estimate the total energy $W_{B}$ contained in the magnetic field by assuming a volume filling factor of unity, which represents an upper limit. For a magnetic field strength of $\sim 30~\mu\mathrm{G}$ and an SNR radius of $20$~pc, we obtain $W_{B} \sim 3.3 \times 10^{49}$~erg, a value comparable to the total energy of electrons ($W_{B}/W_{e} \sim 1.7 $; see Table~\ref{tab:Spec_model_cxou}). This result suggests a near energy equipartition between the magnetic field and the energetic electrons in the emission region, consistent with the expected evolution of young to
middle-aged SNRs \citep{2013ApJ...773..138Y, 2018ApJ...855...59U}. A more precise equipartition could be achieved by reducing the volume filling factor to approximately $0.6$, ensuring $W_{B}/W_{e} \approx 1.0$.

Nevertheless, the total energy inferred for the proton population, $W_{\rm p} = 1.22 \times 10^{51}$~erg, is remarkably high when compared to the typical kinetic energy of a core-collapse supernova ($\sim 10^{51}\text{ erg}$). Such elevated energy levels may suggest that CTB~37B originated from a massive progenitor leading to a superluminous supernova, or from a magnetar-powered explosion, both of which can produce a higher kinetic energy output than a typical supernova \citep[see e.g.][]{1998Natur.395..670G, 2008ASPC..385..109K, 2010ApJ...717..245K, 2012ApJ...746...40L, 2017ApJ...850..148W, 2020NatAs...4..893N}.

In the magnetar-powered supernova scenario, the luminosity and energetics of the explosion are enhanced by the rapid loss of rotational energy from a newly born millisecond magnetar. Through magnetic spin-down, the magnetar injects a large fraction of its rotational energy into the ejecta in the form of a relativistic wind, accelerating and heating the expanding material. This additional energy reservoir can increase the radiated output by one to two orders of magnitude relative to typical core-collapse events, potentially resulting in a superluminous supernova and providing a natural explanation for the large proton energy inferred in our modeling \citep{2010ApJ...717..245K, 2020ApJ...897..152C, 2025A&A...698A..78R}. Nevertheless, it is worth noting that this energy estimate scales inversely with the square of the source distance and is therefore sensitive to uncertainties in the distance determination, which remains poorly constrained.

Another plausible explanation for the observed gamma-ray emission involves the surrounding gas density. A denser medium would reduce the total proton energy required to reproduce the observed gamma-ray flux. Unlike CTB~37A, which exhibits multiple OH (1720~MHz) masers and a well-established SNR--MC interaction \citep{1996AJ....111.1651F, 2000ApJ...545..874R}, CTB~37B does not show a widely confirmed OH (1720~MHz) maser signature. Nevertheless, a potential interaction between molecular gas and CTB~37B has been suggested in previous studies \citep{2012MNRAS.421.2593T, 2015IAUGA..2255641J}. If such an interaction is present, the effective gas density encountered by accelerated particles could be significantly higher than that inferred from X-ray observations.

\begin{figure}
    \includegraphics[width=\hsize,clip]{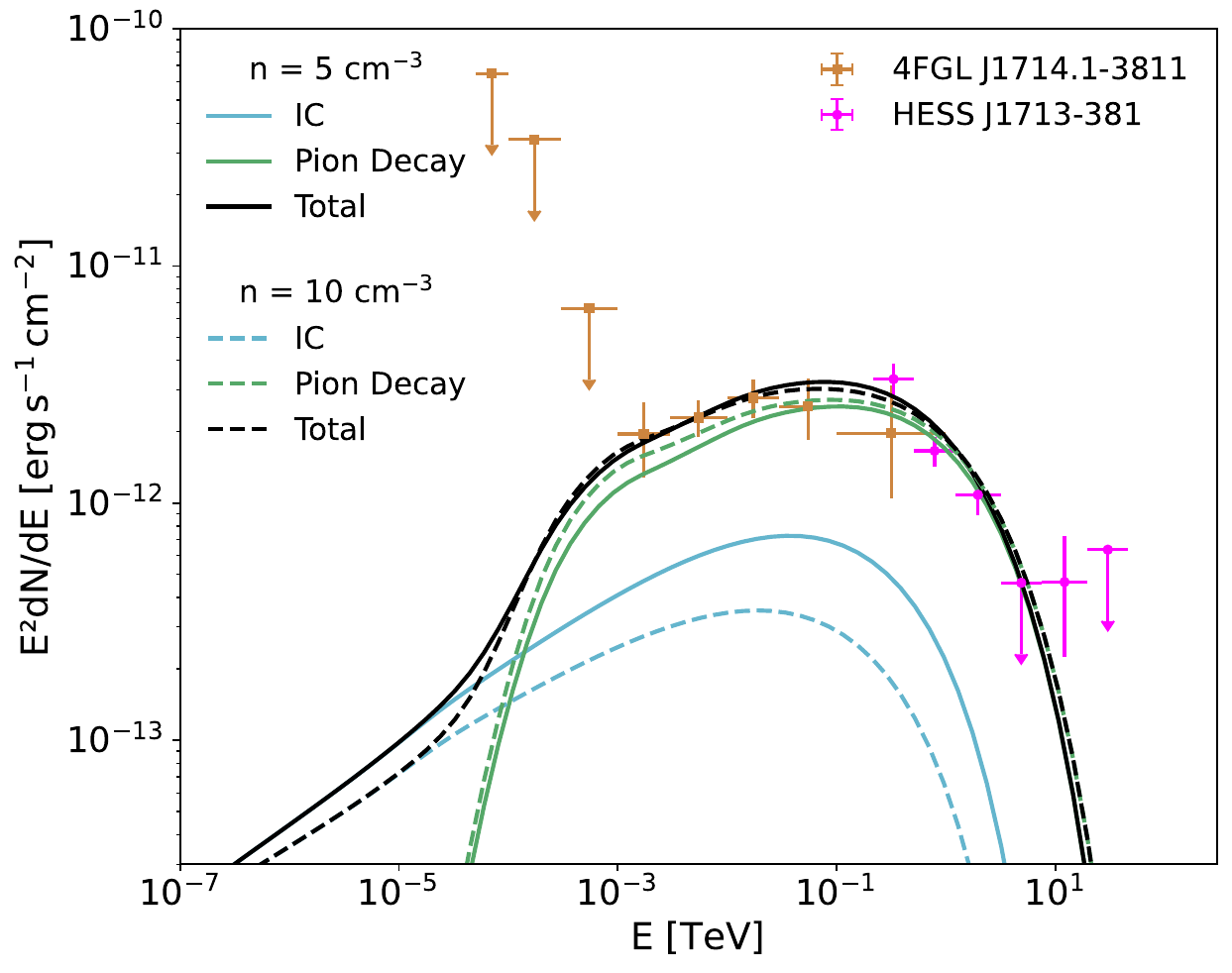}
    \caption{SED of the CXOU~J1714-3810 in the SNR model for the lepto-hadronic-{\it ecpl} model with $K_{ \rm ep} = 10^{-2}$. This model considers a denser environment ($n = 5$~cm$^{-3}$ and $n = 10$~cm$^{-3}$), which is attributed to the SNR's interaction with molecular clouds associated with CTB~37B.}
    \label{fig:cxou_n5_10}
\end{figure}

Figure~\ref{fig:cxou_n5_10} illustrates the impact of adopting effective gas densities of $n = 5$~cm$^{-3}$ and $n = 10$~cm$^{-3}$ on the SED of the SNR region within the lepto-hadronic-\textit{ecpl} model ($K_{\rm ep} = 10^{-2}$). At these higher densities, gamma-ray emission becomes predominantly hadronic, as the increased target gas density enhances pion production. Consequently, the total energy of relativistic protons required to reproduce the observed spectrum decreases to approximately $2.8 \times 10^{50}$~erg for $n = 5$~cm$^{-3}$ and $1.6 \times 10^{50}$~erg for $n = 10$~cm$^{-3}$ \citep[see also][]{2017ApJ...834..153Z}. These values are consistent with the commonly accepted estimate that roughly $10\%$ of the kinetic energy released in a core-collapse supernova is channeled into cosmic-ray production \citep[see e.g.][]{2013A&A...553A..34D, aharonian2013gamma, 2016crpp.book.....G}.

This demonstrates that the gamma-ray flux is sensitive to the surrounding environment, reinforcing the hypothesis that molecular cloud interactions could significantly contribute to the observed emission. These findings highlight the importance of accurately determining the ambient gas density in regions hosting SNR--molecular cloud associations. Future observations of molecular gas tracers, such as CO and H\,{\sc i} line emission, will be essential for refining the gas density estimates and further constraining the role of hadronic processes in the gamma-ray emission from the CXOU~J1714--3810 region \citep{2024A&A...689A.257L, 2024A&A...686A.305F}.

\subsection{MWN scenario}

An alternative explanation for the GeV--TeV emission, explored in Section~\ref{sec:PWN_model_CXOU}, invokes a magnetar-powered PWN. In this scenario, the best fit to the data is achieved with an electron energy distribution, characterized by a spectral index of $\sim 2.5$ and a total energy of $\sim 10^{49}$~erg. This inferred energy substantially exceeds the total energy available from the magnetar's current spin-down power integrated over the lifetime of CTB~37B.

To illustrate this discrepancy, we estimate the current energy budget by integrating the time-dependent spin-down power over the magnetar's lifetime \citep[see e.g.][]{2018A&A...612A...2H, 2014JHEAp...1...31T}. Assuming that the spin-down evolution follows a braking index of $n \sim 3$ and a typical initial spin-down timescale of $\tau_0 \sim 10^{3}$~yr \citep{2012MNRAS.427..415M, 2009ApJ...703.2051G}, and adopting the present spin-down luminosity $\dot{E}_{\rm sd} \approx 4.5 \times 10^{34}$~erg~s$^{-1}$ and the characteristic age $\tau_{\rm c} \approx 0.95$~kyr \citep{2014ApJS..212....6O}, we obtain a total rotational energy release of only $E_{\rm sd} \approx 2.6 \times 10^{45}$~erg -- several orders of magnitude below the total energy required by the relativistic electrons. Even assuming that the magnetar has the same age as the associated SNR, this value increases to $E_{\rm sd} \approx 4.2 \times 10^{46}$~erg, still roughly two orders of magnitude too low. Furthermore, the present spin-down luminosity is at least two orders of magnitude weaker than that of young pulsars powering known TeV PWNe \citep[e.g. Kes~75, Vela~X;][]{2018A&A...612A...2H}.

This energy budget problem can be resolved by considering magnetar formation theories, which posit initial rotation periods on the order of milliseconds to generate the intense magnetic fields via a turbulent dynamo \citep{1993ApJ...408..194T, 2020MNRAS.494.4838L}. For instance, with a short initial period of $P_0 < 0.01$~s, the magnetar's initial rotational energy would exceed $2 \times 10^{50}$~erg assuming the characteristic age, or $10^{51}$~erg if the age of the SNR is adopted. These values are more than sufficient to power the MWN and account for the observed non-thermal emission.

Therefore, inverse-Compton emission from a magnetar-powered PWN is viable, as it would require converting only a small fraction ($<~10^{-2}$--$10^{-3}$) of the magnetar's initial rotational energy into relativistic electrons. However, the MWN scenario faces a significant challenge from the observed morphology.

The TeV source HESS~J1713--381 has an angular size of $0.092^{\circ}$ \citep{2018A&A...612A...1H}, corresponding to a physical radius of $\sim 21$~pc at a distance of 13.2~kpc. In contrast, the GeV source 4FGL~J1714.1--3811 is more compact ($0.027^{\circ}$; \citealt{2022ApJS..260...53A}), with a radius of $\sim 6$~pc. Assuming a characteristic magnetar age of $\sim 1$~kyr and a magnetic field strength of $B = 5.18$--$11.3\,\mu$G (see Table~\ref{tab:Spec_model_cxou}), the theoretical diffusion length scale \citep[$R_{\rm diff}$; see][]{2023ApJ...945...66P} of the TeV-emitting electrons within an associated MWN is constrained to just $\sim 4$--$6$~pc. Even adopting the older dynamical age of the supernova remnant, the MWN would expand to a maximum radius of only $\sim 9$--$13$~pc. These physical scales are marginally consistent with the GeV source extent, but they cannot account for the much larger TeV emission region. This spatial mismatch strongly disfavors a dominant MWN contribution and instead favors the SNR scenario, where the extended TeV morphology naturally aligns with the known radio size of CTB~37B.

Consequently, the supernova remnant scenario emerges as the most plausible framework for explaining the high- and very-high-energy emission from this region. Because the purely leptonic model underestimates the \hess energy flux around $\sim 10$~TeV, a lepto-hadronic model provides a more satisfactory description of the GeV--TeV data. This interpretation, however, requires either that CTB~37B originated from an unusually energetic explosion, such as a magnetar-powered supernova, or that the ambient medium is denser than inferred from current X-ray measurements.

\section{CTAO Contributions} \label{sec:CTAO}

\begin{figure*} 
    \centering
    \begin{subfigure}{0.49\textwidth}
        \includegraphics[width=\linewidth]{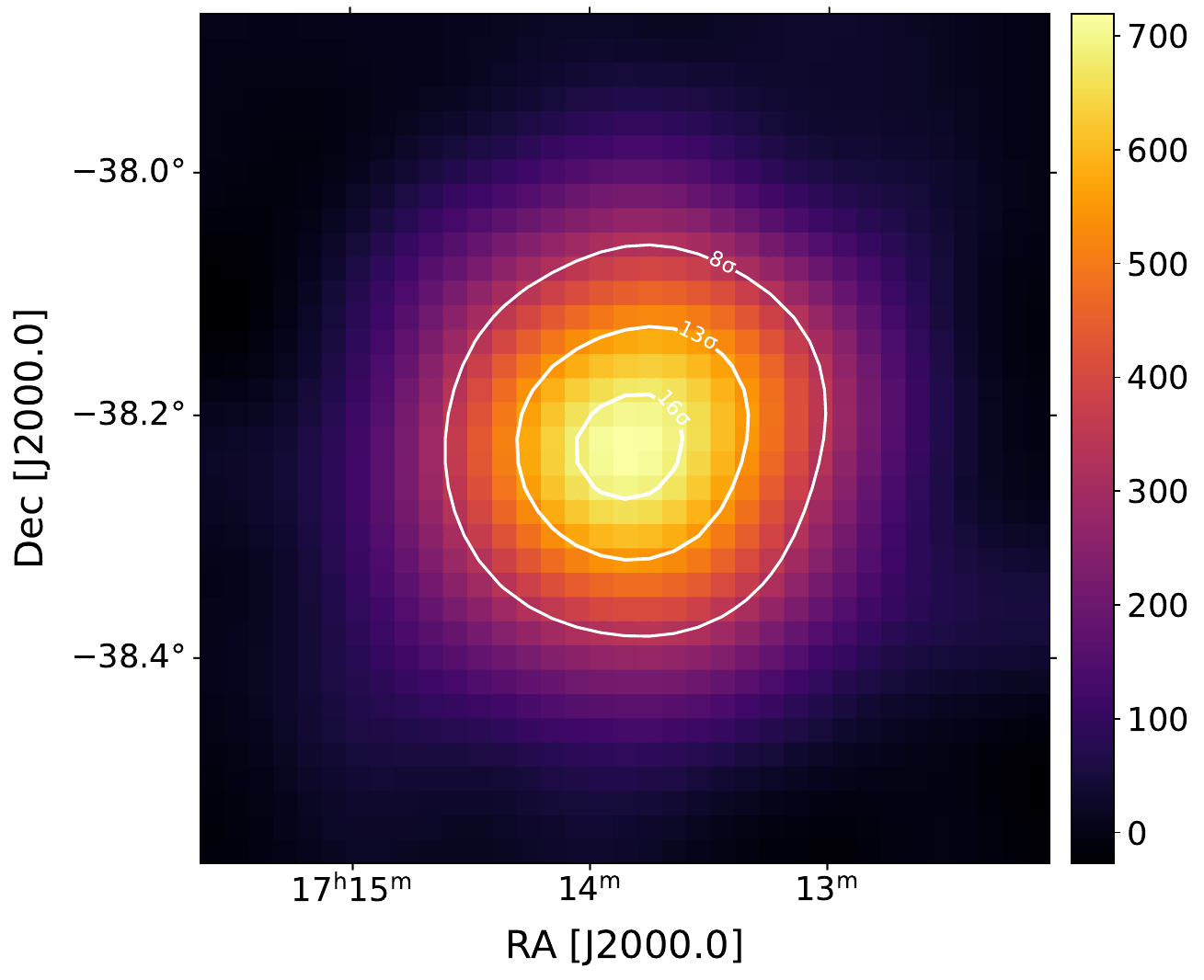}
    \end{subfigure}
    \hfill
    \begin{subfigure}{0.49\textwidth}
        \includegraphics[width=\linewidth]{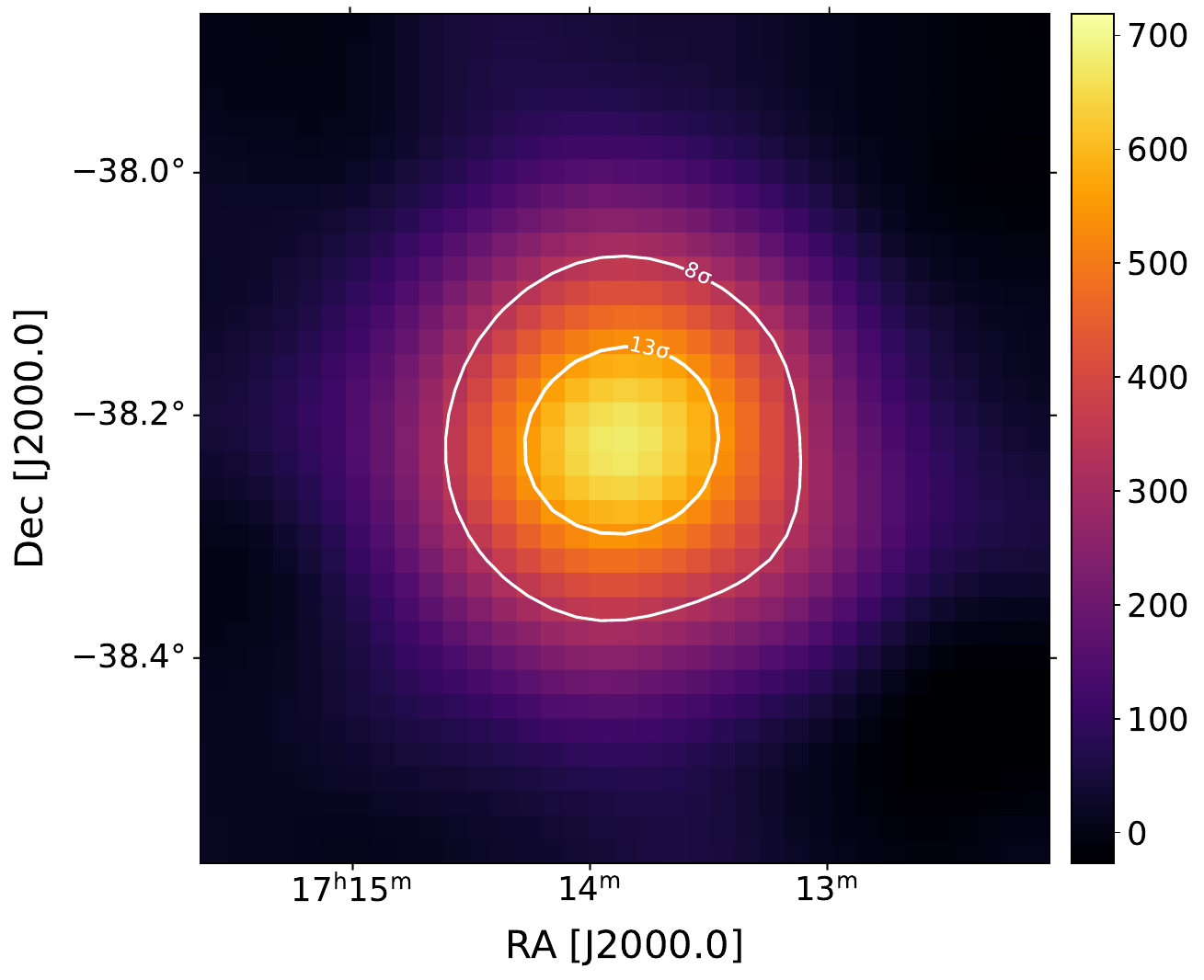}
    \end{subfigure}
    \hfill
    \begin{subfigure}{0.49\textwidth}
        \includegraphics[width=\linewidth]{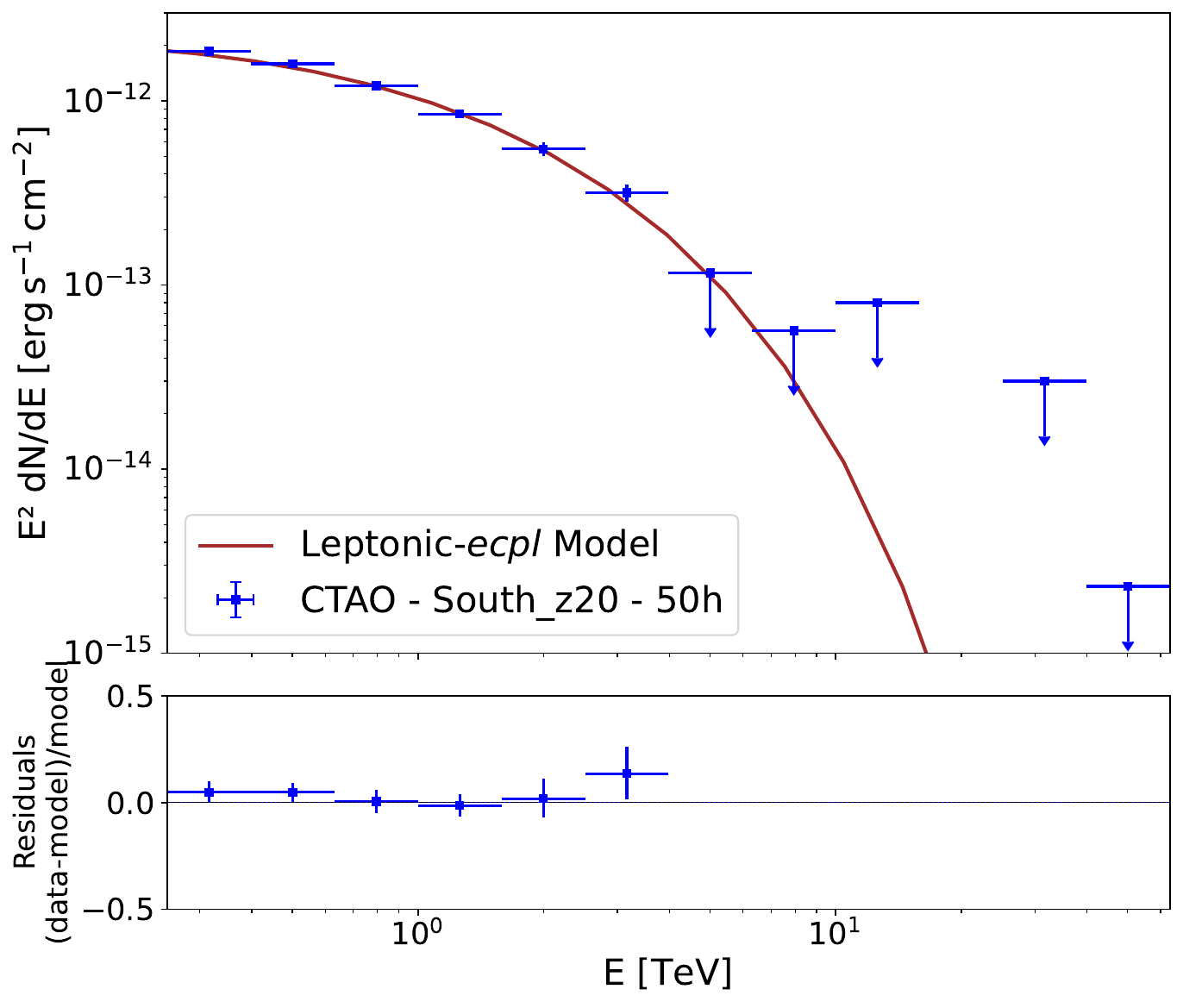}
        \caption{Leptonic-{\it ecpl} model}
    \end{subfigure}
    \hfill
    \begin{subfigure}{0.49\textwidth}
        \includegraphics[width=\linewidth]{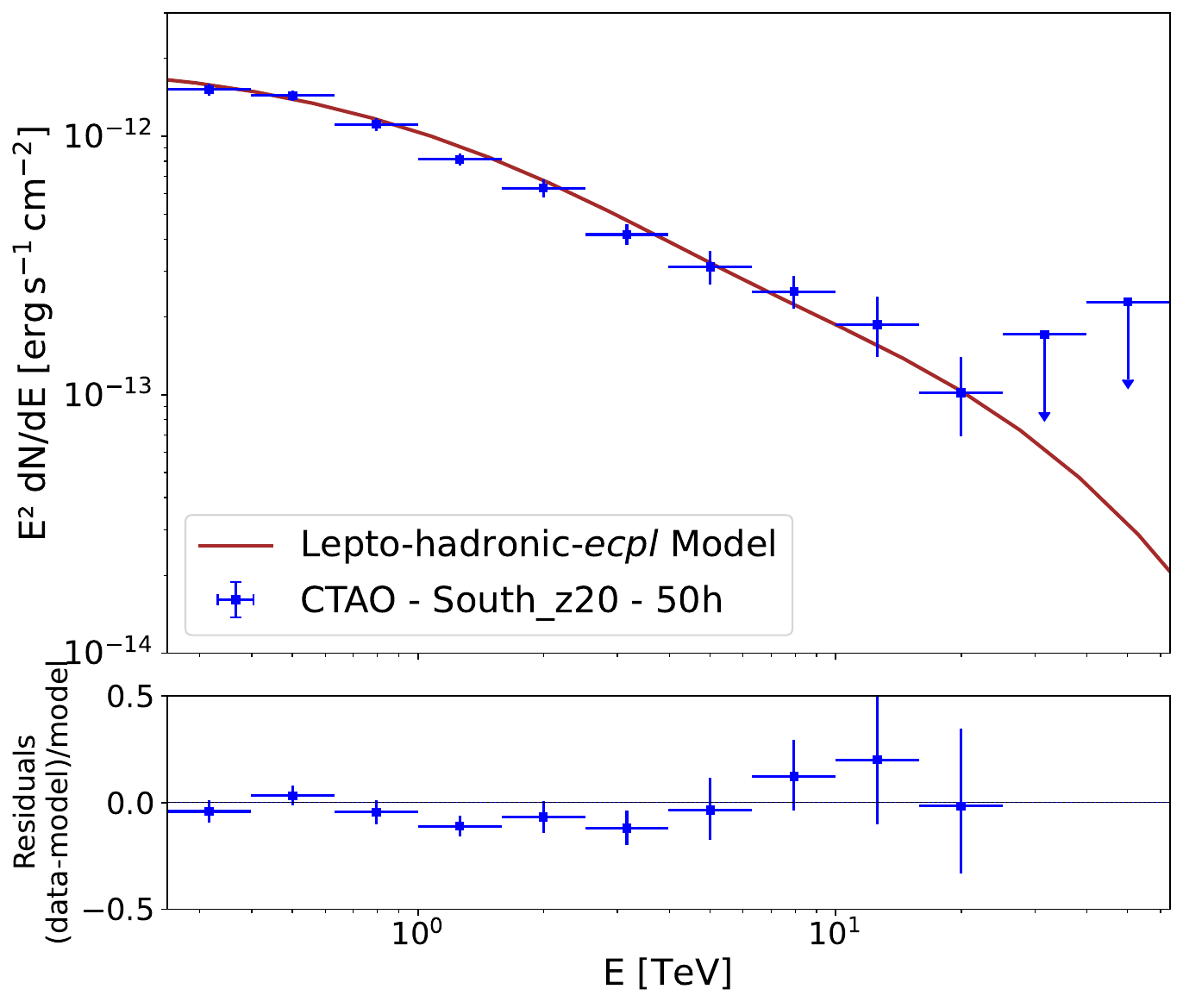}
        \caption{Lepto-hadronic-{\it ecpl} model}
    \end{subfigure}
    \caption{Simulated observations of CTAO for CTB 37B/CXOU J1714-3810 region assuming an exposure time of $t_{\mathrm{obs}} = 50$ h, employing the full southern-array IRFs at a zenith angle of $z = 20^{\circ}$. Upper panels: Simulated VHE gamma-ray excess maps for the leptonic (left) and lepto-hadronic (right) emission scenarios. The maps are generated with a spatial binning of $0.02^{\circ}$ and smoothed using a Gaussian kernel with width $0.05^{\circ}$. The solid white contours correspond to detection significance levels of $8\sigma$, $12\sigma$, and $16\sigma$. Lower panels: Reconstructed spectral flux points overlaid on the assumed broadband SED models for each emission scenario. The close agreement between the reconstructed spectra and the input models highlights the capability of CTAO to recover the intrinsic source spectrum with high accuracy.}
    \label{fig:fp_ctao}
\end{figure*}

The Cherenkov Telescope Array Observatory \citep[CTAO;][]{2019scta.book.....C} is the next-generation ground-based facility for very-high-energy gamma-ray astronomy, designed to probe non-thermal emission from astrophysical sources. As an open observatory, CTAO will operate from two sites, one in each hemisphere, ensuring full-sky coverage and achieving an order-of-magnitude improvement in sensitivity over current instruments \citep{2019scta.book.....C}. In addition, its angular resolution will substantially surpass that of wide-field air-shower observatories such as HAWC and LHAASO. These capabilities make CTAO particularly well suited for detailed investigations of the spectral and morphological characteristics of the region surrounding CXOU J1714-3810. Moreover, this source lies within the observational sky coverage of CTAO South, rendering our simulations directly relevant for forthcoming survey and targeted observation studies.

To assess the capabilities of the CTAO in the CXOU~J1714$-$3810 region and determine its ability to differentiate between the leptonic and lepto-hadronic scenarios, we carried out dedicated three-dimensional map simulations and derived the corresponding spectral flux points. Prior to these simulations, we estimated the annual visibility of the source position for both CTAO South and CTAO North during the year 2026, adopting zenith-angle ($z$) constraints of $20^\circ$, $40^\circ$, and $60^\circ$. For CTAO South, the source is observable under all three zenith-angle selections, with total visibility times of approximately $793.9$ h ($z=20^\circ$), $644.8$ h ($z=40^\circ$), and $642.5$ h ($z=60^\circ$). In contrast, CTAO North can access the region only at large zenith angles ($z=60^\circ$), yielding a visibility of approximately $548.5$ h. Because observations performed at high zenith angles are affected by stronger atmospheric absorption and reduced angular resolution, we adopted the full CTAO South array configuration at $z=20^\circ$, which provides the longest effective exposure together with near-optimal instrumental performance and favorable observing conditions. All subsequent simulations were therefore carried out using the \texttt{prod5}~v0.1 instrument response functions (IRFs) for CTAO South at $z=20^{\circ}$. These IRFs correspond to the CTAO alpha configuration, representing the initial deployment phase of the observatory \citep{cherenkov_telescope_array_observatory_2021_5499840}.

The simulated very-high-energy gamma-ray excess maps and reconstructed spectral flux points were produced using the \texttt{Gammapy} software \citep{2023A&A...678A.157D, acero_2025_17814297}, following a standard 3D likelihood analysis for an exposure time of $t_{\rm obs}=50$~h. We generated a counts cube with a spatial binning of $0.02^{\circ}$ and 10 logarithmically spaced energy bins per decade covering the energy range from $100$~GeV to $70$~TeV, consistent with the nominal sensitivity of the CTAO southern site.

The emission from the region was modeled within the supernova remnant scenario using the leptonic-\textit{ecpl} and lepto-hadronic-\textit{ecpl} particle-distribution models. These models were selected based on their statistical preference in the spectral fitting procedure, as well as on their physical consistency within the proposed interpretation. In both scenarios, the SNR emission was described using a Gaussian spatial template (see Table~\ref{tab:Spatial_model}), with parameters adopted from the morphological analyses reported for HESS~J1713-381 \citep[see][]{2018A&A...612A...1H}. For each emission scenario, we produced 3000 independent realizations of the simulated observations by applying Poisson fluctuations to the expected source and background event distributions, including the background component derived from the IRFs. The reconstructed spectral flux points and significance maps were then obtained through a maximization of the binned Poisson likelihood. Figure~\ref{fig:fp_ctao} presents the results for leptonic-\textit{ecpl} and lepto-hadronic-\textit{ecpl}. The contour levels (solid white lines) shown in the maps correspond to significances defined as $\sigma = \sqrt{\rm TS}$, where TS represents the test statistic.

\setlength{\tabcolsep}{30pt}
\begin{table}
\centering
\renewcommand{\arraystretch}{1.3}
\caption{Spatial model employed in the 3D simulations of CTAO observations. The adopted morphological parameters are derived from H.E.S.S. analysis of HESS~J1713-381, whose measurements span an energy range comparable to that considered in the present CTAO simulations \citep{2018A&A...612A...1H}.}
\label{tab:Spatial_model}
\begin{tabular}{l c}
\hline
\hline
\textbf{Parameter} & \textbf{Value} \\
\hline
Spatial model & 2D Gaussian \\
RA (J2000.0) ($^\circ$) & $258.46 \pm 0.015$ \\
Dec (J2000.0) ($^\circ$) & $-38.22 \pm 0.014$ \\
$1\sigma$ extension ($^\circ$) & $0.092 \pm 0.017$ \\
\hline
\hline
\end{tabular}
\vspace{0.2cm}
\end{table}

For an exposure time of $50$ h, the source is detected with high statistical significance, with contours in the sky maps reaching levels of approximately $8\sigma$, $13\sigma$, and $16\sigma$. These results demonstrate that the CTAO will be capable of spatially resolving the SNR with high confidence. In the lepto-hadronic scenario, the emission appears more spatially diffuse and exhibits a lower peak $\gamma$-ray excess. By contrast, the leptonic model produces a more centrally concentrated morphology, resulting in enhanced significance levels toward the central region of the source. This behaviour is also evident in the overlaid significance contours, which display a more pronounced central enhancement in the leptonic case.

It is worth mentioning that these morphological differences are not intrinsic to the simulated source profiles. Indeed, both scenarios use the same 2D Gaussian spatial template under the standard assumption of spatial-spectral factorization in the \texttt{Gammapy} framework, where the total source flux is modelled as the product of independent spatial and spectral functions. Instead, this behavior is an instrument-induced effect driven by how the differing spectral shapes interact with the energy-dependent IRFs and local detection thresholds of the CTAO. In the lepto-hadronic framework, the harder compound spectrum extends to multi-TeV energies, a regime where the CTAO exhibits a larger effective area and a suppressed cosmic-ray background. This elevated signal-to-noise ratio allows the faint outer wings of the intrinsic Gaussian profile to emerge above the statistical background, producing a more extended appearance in the reconstructed morphology. Conversely, in the leptonic scenario, the flux drops exponentially around $\sim 5$~TeV, where the instrumental background is higher, so the outer edges of the spatial distribution fall below the detection threshold. Only the bright central core remains detectable, making the source appear more concentrated.

The reconstructed spectral flux points closely reproduce the input spectral models, with relative residuals ($(\mathrm{data}-\mathrm{model})/\mathrm{model}$) remaining below $0.5$ over the full explored energy range. This indicates that CTAO will be able to recover the intrinsic spectral shape of the source with high reliability. More importantly, the two emission scenarios predict markedly different behaviours at the highest energies. In the leptonic case, the spectrum steepens rapidly above $\sim 5$~TeV as a consequence of the maximum electron energy constrained by the upper limits on the synchrotron X-ray emission. In contrast, the lepto-hadronic scenario preserves a significantly harder, power-law-like spectral tail extending up to $\sim 50$~TeV, driven by sustained proton--proton ($pp$) interactions. The simulated flux points clearly trace these differences in the highest-energy bins, demonstrating that CTAO observations will be able to discriminate between the two emission scenarios on spectral grounds.

\begin{figure*} 
    \centering
    \begin{subfigure}{0.49\textwidth}
        \includegraphics[width=\linewidth]{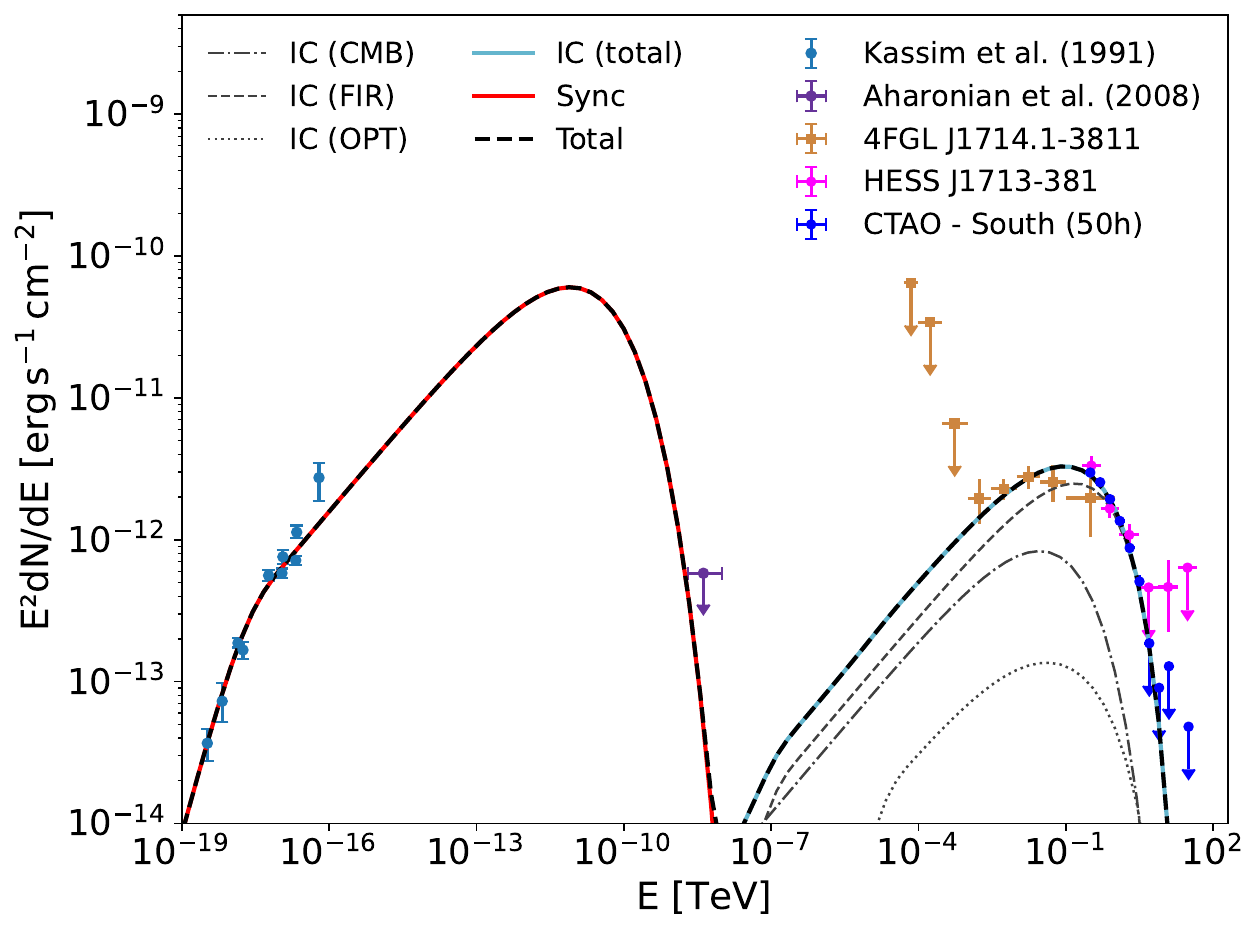}
        \caption{Leptonic-{\it ecpl} model}
    \end{subfigure}
    \hfill
    \begin{subfigure}{0.49\textwidth}
        \includegraphics[width=\linewidth]{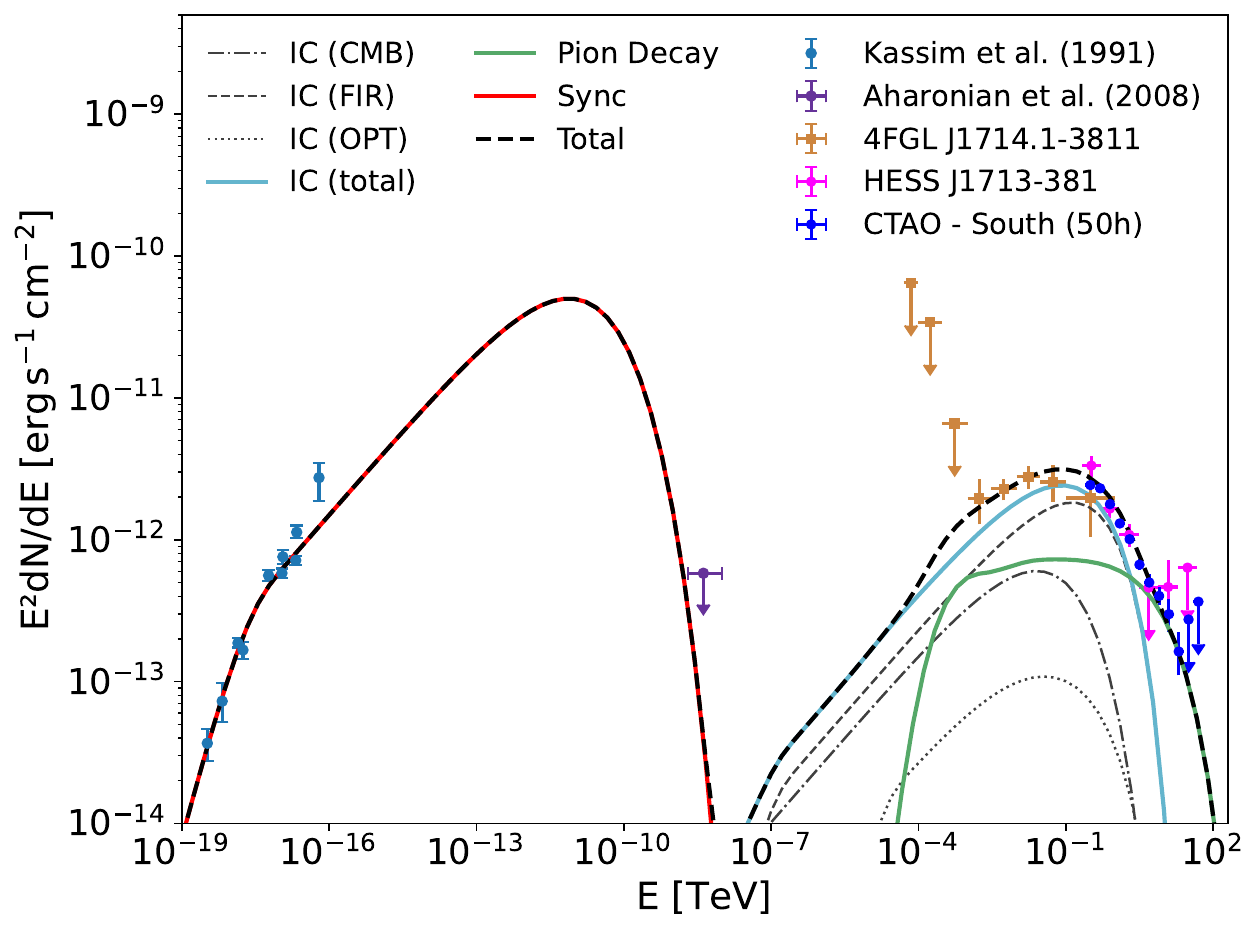}
        \caption{Lepto-hadronic-{\it ecpl} model}
    \end{subfigure}
    \caption{SED of the region associated with CTB 37B/CXOU~J1714-3810, including simulated flux points for the CTAO assuming an observation time of $t_{\rm obs}=50$ h. The emission is modeled using a leptonic-{\it ecpl} scenario (left panel) and a lepto-hadronic-{\it ecpl} scenario with $K_{\rm ep}=10^{-2}$ (right panel), both of which provide statistically favored representations of the observational data.}
    \label{fig:cxou_ctao}
\end{figure*}

Quantitatively, the expected detection significance is $\sim 27\sigma$ for the leptonic model and $\sim 26\sigma$ for the lepto-hadronic model, with the slight difference arising from their distinct spectral shapes. For comparison with previous observations, Fig.~\ref{fig:cxou_ctao} displays the simulated CTAO flux points alongside existing data and the best-fitting multiwavelength SED. The energy flux uncertainties for CTAO are remarkably small, particularly at lower energies, confirming that CTAO will probe this regime with significantly greater precision than current observatories such as \hess. While the CTAO results for the leptonic model are consistent with existing \hess measurements, CTAO will improve upon the \hess upper limits by at least a factor of 3 at energies above $5\text{ TeV}$. For the lepto-hadronic model, the simulated spectrum extends beyond $\sim 5\text{ TeV}$, a regime where there are currently upper limits measured by \hess. Consequently, a 50-h observation will provide much tighter constraints on the cutoff energies of both the electron and proton populations.

In summary, CTAO observations of this magnetar region will provide critical insight into the energy-dependent behavior of the TeV emission, enabling more precise constraints on the maximum particle energies responsible for the observed spectra.

\section{Conclusions} \label{sec:conc}

We have investigated the non-thermal emission from the vicinity of the magnetar CXOU~J1714--3810 by modeling its multiwavelength spectral energy distribution under leptonic and lepto-hadronic scenarios. Using MCMC sampling implemented in the \texttt{Naima} software, we constrained the underlying particle populations responsible for the observed radio-to-TeV emission, disentangling the relative contributions of synchrotron radiation, inverse-Compton scattering, and neutral-pion decay.

Within the SNR interpretation, both purely leptonic and lepto-hadronic models provide acceptable fits to the current GeV--TeV data. The leptonic scenario, characterized by an \textit{ecpl} electron distribution with a cutoff at a few TeV and a magnetic field strength of $B \sim 28~\mu$G, reproduces the overall spectral shape but slightly underestimates the highest-energy H.E.S.S.\ flux point. By contrast, the lepto-hadronic model with an electron-to-proton ratio $K_{\rm ep} \sim 10^{-2}$ yields a more complete description of the spectrum, particularly above $\sim 10$~TeV. In this scenario, IC scattering dominates the GeV emission, while pion decay becomes increasingly important at the highest energies. The inferred total energy in relativistic protons, $W_{\rm p} \gtrsim 10^{51}$~erg, is comparable to or exceeds the canonical kinetic energy of a core-collapse supernova. This apparent tension can be mitigated if CTB~37B resulted from a more energetic explosion, such as a magnetar-powered or superluminous supernova, or if the remnant is interacting with a denser ambient medium. For effective gas densities of $n \sim 5$--$10$~cm$^{-3}$, the required proton energy decreases to values consistent with a $\sim 10\%$ conversion of the supernova kinetic energy into cosmic rays.

We also considered an MWN scenario in which the gamma-ray emission is powered by the rotational energy of CXOU~J1714--3810. While inverse-Compton emission from a magnetar-powered nebula can reproduce the broadband spectral shape and may be energetically viable if the magnetar was born with a millisecond spin period, this interpretation is strongly disfavored by morphological constraints. In particular, the large spatial extent of the TeV emission is incompatible with the expected size of an MWN at the inferred age of the system, whereas it is naturally explained by the known radio size of CTB~37B. This spatial mismatch strongly favors an SNR origin for the TeV emission. Nevertheless, a contribution from an MWN cannot be entirely excluded. If present, such emission would likely be confined to an angular size of only $\sim 0.01^\circ - 0.04^\circ$ around the magnetar position and would have a flux smaller than, or at most comparable to, that of the SNR component, given that the current gamma-ray data are already well reproduced within the SNR scenario.

Finally, we assessed the prospects for future observations with CTAO. Our simulations indicate that CTAO will provide a powerful test of the emission mechanisms operating in the vicinity of CXOU~J1714-3810. For an exposure of 50 h, CTAO will detect the source with high statistical significance and spatially resolve the SNR morphology with high confidence. The instrument will also accurately recover the intrinsic spectral shape of the source over a broad energy range. Most importantly, the leptonic and lepto-hadronic scenarios predict markedly different behaviors above a few TeV: the leptonic model exhibits a rapid spectral steepening due to the limited maximum energy of the accelerated electrons, whereas the lepto-hadronic model maintains a hard spectral tail extending up to tens of TeV as a result.

We note that the stationary one-zone SED models employed here are simplified descriptions of intrinsically complex environments. Time evolution can alter particle distributions and the resulting spectra, and spatial complexity is expected when X-ray and TeV emission regions differ in size or morphology \citep[e.g.][]{2011ApJ...742...62V, 2013ApJ...773..139V, 2017ApJ...835...54T}. The inferred parameters are also sensitive to environmental inputs such as ambient density and the energy densities of the infrared and optical photon fields. As a result, the MCMC posteriors exhibit parameter covariances, and the quoted values should be regarded as phenomenological constraints that compare the relative plausibility of emission scenarios rather than as precise physical measurements. In addition, the results depend on the adopted upper limits for the non-thermal X-ray emission. More restrictive X-ray constraints would reduce the allowed synchrotron flux and, consequently, require lower electron energy densities and/or lower maximum particle energies. Such conditions would tend to favor models with smaller $K_{\rm ep}$ values in the lepto-hadronic scenario and could shift the preferred spectral solutions toward lower cutoff energies.

In conclusion, our results suggest that the high-energy emission observed in regions hosting magnetars can be influenced by the interplay between supernova remnants, molecular clouds, and compact objects. The gamma-ray radiation may originate from hadronic interactions associated with an SNR–molecular cloud system, while the presence of a magnetar can further affect the energy budget through its rotational energy losses and magnetic activity, particularly during the early stages of its evolution. Future CTAO observations will be crucial for testing these scenarios, offering new insights into the role of magnetars, their wind nebulae, and their surrounding environments as particle accelerators and sources of high-energy radiation.

\section*{Acknowledgements}

We sincerely thank the referee for their thoughtful feedback and valuable suggestions, which have greatly enhanced the clarity and scientific rigor of this work. M.F.S. thanks a Fundação de Amparo à Pesquisa do Estado de São Paulo (FAPESP, grants No. 2025/05794-2 and No. 2021/01089-1) for the financial support. He also thanks the support of Conselho Nacional de Desenvolvimento Cient\'{i}fico e Tecnol\'{o}gico (CNPq, No. 173535/2023-2). R.C.A. acknowledges the financial support of the NAPI “Fenômenos Extremos do Universo” of Fundação de Apoio à Ciência, Tecnologia e Inovação do Paraná. R.C.A. research is supported by CAPES/Alexander von Humboldt Program (grant No. 88881.800216/2022-01), CNPq (grant Nos. 310448/2021-2 and 4000045/2023-0), Araucária Foundation (grant Nos. 698/2022 and 721/2022) and FAPESP (grant No. 2021/01089-1). R.C.A. also acknowledges the support of L’Oreal Brazil, with the partnership of ABC and UNESCO in Brazil. The authors acknowledge the AWS Cloud Credit/CNPq and the National Laboratory for Scientific Computing (LNCC/MCTI, Brazil) for providing HPC resources of the SDumont supercomputer, which have contributed to the research results reported in this paper. URL: https://sdumont.lncc.br. The research also used Gammapy, a Python package developed by the community for TeV gamma-ray astronomy \citep{Deil_2017, 2023A&A...678A.157D}, accessible at \href{https://www.gammapy.org}{https://www.gammapy.org}.

\section*{Data Availability}

The data underlying this article will be shared on reasonable request to the corresponding author. In addition, we used the instrument response functions for the Cherenkov Telescope Array Observatory (CTAO) provided by the CTA Consortium and CTAO. For detailed information on these instrument response functions, see \href{https://www.ctao-observatory.org/science/cta-performance}{https://www.ctao-observatory.org/science/cta-performance} \citep[version prod5 v0.1;][]{cherenkov_telescope_array_observatory_2021_5499840}.



\bibliographystyle{mnras}
\bibliography{references} 








\bsp	
\label{lastpage}
\end{document}